\documentclass[12pt]{article}     
\usepackage[utf8]{inputenc}
\usepackage[russian]{babel}
\usepackage{graphicx}
\usepackage{verbatim}
\usepackage{amsmath}
\usepackage [T2A]{fontenc}
\usepackage {amsthm}
\usepackage {amssymb}
\usepackage {float}
\usepackage {textcomp}
\usepackage {euscript}
\usepackage [nottoc]{tocbibind}
\usepackage {listings}
\usepackage{multirow}
\usepackage{hhline}
\newtheorem{theorem}{Теорема}[subsection]
\newtheorem{lemma}[theorem]{Лемма}

\newtheorem{statement}[theorem]{Утверждение}
\newtheorem*{follow}{Следствие}
\newtheorem*{note}{Замечание}
\sloppypar

\begin{document}
\title{Критерий неявной полноты в трехзначной логике в 
терминах предполных классов}
\author{М.\,В.~Старостин
\thanks{{\it Старостин Михаил Васильевич } --- асс. каф. дискретной математики мех.-мат. ф-та МГУ,
e-mail: murmol@bk.ru}}
\date{Март 2021}
\maketitle
\begin{abstract}
    Понятие неявной выразимости было введено А.\,В.~Кузнецовым 
    в 1979 году как обобщение выразимости суперпозициями. 
    Система функций называется неявно полной, 
    если всякая функция имеет 
    неявное представление над этой системой.
    В настоящей работе был получен критерий неявной полноты
    в трехзначной логике $P_3$ в терминах 
    максимальных неполных (предполных) классов.
    \medskip
    
    {\it Ключевые слова}: трехзначная логика, 
     неявная выразимость, предполные классы.
\end{abstract}

\newcommand{\ufunc}[3]
{
\boxed{
\begin{matrix}
#1\\#2\\#3
\end{matrix}
}
}

\newcommand{\sooo}{\ufunc{0}{0}{0}}
\newcommand{\sooi}{\ufunc{0}{0}{1}}
\newcommand{\sooz}{\ufunc{0}{0}{2}}
\newcommand{\soio}{\ufunc{0}{1}{0}}
\newcommand{\soii}{\ufunc{0}{1}{1}}
\newcommand{\soiz}{\ufunc{0}{1}{2}}
\newcommand{\sozo}{\ufunc{0}{2}{0}}
\newcommand{\sozi}{\ufunc{0}{2}{1}}
\newcommand{\sozz}{\ufunc{0}{2}{2}}
\newcommand{\sioo}{\ufunc{1}{0}{0}}
\newcommand{\sioi}{\ufunc{1}{0}{1}}
\newcommand{\sioz}{\ufunc{1}{0}{2}}
\newcommand{\siio}{\ufunc{1}{1}{0}}
\newcommand{\siii}{\ufunc{1}{1}{1}}
\newcommand{\siiz}{\ufunc{1}{1}{2}}
\newcommand{\sizo}{\ufunc{1}{2}{0}}
\newcommand{\sizi}{\ufunc{1}{2}{1}}
\newcommand{\sizz}{\ufunc{1}{2}{2}}
\newcommand{\szoo}{\ufunc{2}{0}{0}}
\newcommand{\szoi}{\ufunc{2}{0}{1}}
\newcommand{\szoz}{\ufunc{2}{0}{2}}
\newcommand{\szio}{\ufunc{2}{1}{0}}
\newcommand{\szii}{\ufunc{2}{1}{1}}
\newcommand{\sziz}{\ufunc{2}{1}{2}}
\newcommand{\szzo}{\ufunc{2}{2}{0}}
\newcommand{\szzi}{\ufunc{2}{2}{1}}
\newcommand{\szzz}{\ufunc{2}{2}{2}}

\section{Введение}
\addtocounter{subsection}{1}
Обозначим множество $\{0,1,\dots,k-1\}$ через $E_k$,
а множество всех функций, определенных на $E_k$ через $P_k$.

Пусть $\mathfrak{A}$~--- система функций $k-$значной логики.
\textit{Явным замыканием} $\mathfrak{A}$ будем называть
систему $E(\mathfrak{A}) = [\mathfrak{A}\cup \{x\}]$, 
где $[\cdot]$ обозначает замыкание по суперпозиции.

Рассмотрим систему уравнений
$$
\left\{
\begin{array}{ccl}
A_1 (x_1,\dots,x_n, y_1,\dots, y_p, z)=
B_1 (x_1,\dots,x_n, y_1,\dots, y_p, z),\\
A_2 (x_1,\dots,x_n, y_1,\dots, y_p, z)=
B_2 (x_1,\dots,x_n, y_1,\dots, y_p, z),\\
\dots\\
A_m (x_1,\dots,x_n, y_1,\dots, y_p, z)=
B_m (x_1,\dots,x_n, y_1,\dots, y_p, z),\\
\end{array}
\right.
$$
где $A_1,\dots,A_m,B_1,\dots,B_m$ --- функции из $E(\mathfrak{A})$.
Переменные $x_1,\dots,x_n$ 
будем называть \textit{основными} переменными,
переменные $y_1,\dots,y_p$ --- \textit{параметрами},
а $z$ --- \textit{выделенной} переменной.
Говорят, что данная система является 
\textit{параметрическим представлением} 
функции $f(x_1,\dots,x_n)\in P_k$, 
если она имеет хотя бы одно решение вида
$$
\begin{array}{ccl}
y_1 = g_1(x_1,\dots,x_n),\\
\dots\\
y_p = g_p(x_1,\dots,x_n),\\
z = g_0(x_1,\dots,x_n),
\end{array}
$$
причем для каждого решения такого вида выполнено равенство 
$$
g_0(x_1,\dots,x_n) = f(x_1,\dots,x_n).
$$

Функция называется \textit{параметрически выразимой} 
над системой функций $\mathfrak{A}$,
если для нее существует 
параметрическое представление над $\mathfrak{A}$.
Если для функции $f(x_1,\dots,x_n)$ 
существует параметрическое представление над $\mathfrak{A}$,
в котором $p=0$ (то есть параметры отсутствуют), 
то соответствующая система уравнений называется 
\textit{неявным представлением} функции $f$,
а функция $f$ называется \textit{неявно выразимой} 
над системой функций $\mathfrak{A}$.
Подробнее об этом см.~\cite{Kuznetsov, OMKZ_bf_param}.

Множество всех функций, 
параметрически выразимых над системой функций $\mathfrak{A}$
называется \textit{параметрическим замыканием} $\mathfrak{A}$ и
обозначается через $P(\mathfrak{A})$.
Система $\mathfrak{A}$ называется \textit{параметрически замкнутой},
если она совпадает со своим параметрическим замыканием.
Множество всех функций, 
неявно выразимых над системой функций $\mathfrak{A}$
называется \textit{неявным расширением} $\mathfrak{A}$ и
обозначается через $I(\mathfrak{A})$.
Легко видеть, что
$\mathfrak{A}\subseteq [\mathfrak{A}] \subseteq E(\mathfrak{A})
\subseteq I(\mathfrak{A}) = I(E(\mathfrak{A}))
\subseteq P(\mathfrak{A})$.

Оператор параметрического замыкания является оператором 
замыкания в общем смысле,
то есть обладает свойствами 
экстенсивности, изотонности и идемпотентности.
Кузнецов в работе~\cite{Kuznetsov} доказал, 
что в $P_2$ имеется 25 параметрически замкнутых классов и 
перечислил их все.
Оператор неявного расширения в общем случае 
не обладает свойством идемпотентности, однако,
как показал О.М. Касим-Заде в~\cite{OMKZ_P2}, в двузначной логике
неявная выразимость эквивалентна параметрическому замыканию и,
следовательно, оператор неявной выразимости в $P_2$~--- 
оператор замыкания.

Система $\mathfrak{A}$ называется 
\textit{параметрически (неявно) полной в $P_k$}, 
если $P(\mathfrak{A})$ (соответсвенно $I(\mathfrak{A})$) 
совпадает с $P_k$.
Система $\mathfrak{A}$ называется 
\textit{параметрически (неявно) предполной в $P_k$}, 
если она не является параматрически (неявно) полной,
но для любой функции $f\notin \mathfrak{A}$
система $\mathfrak{A}\cup \{f\}$ параметрически (неявно) полна. 

В $P_2$ имеется 6 параметрически (и, соответственно, неявно)
предполных классов~\cite{Kuznetsov, OMKZ_P2}: 
$T_0, T_1, L, S, K, D$ 
(здесь и далее обозначения замкнутых классов в $P_2$ 
даны по~\cite{Ugolnikov}).
Кроме того, в $P_2$ имеется один 
минимальный (по включению) замкнутый по суперпозиции 
неявно полный класс $M$~\cite{OMKZ_crypto}.
В дальнейшем такие классы мы будем называть
минимальными неявно полными классами.

В трехзначной логике все минимальные неявно полные классы описаны 
Е.А. Ореховой в~\cite{Orehova_P3}. 
Каждый из них двойственен относительно некоторой подстановки 
замыканию по суперпозиции одной из следующих систем:
\begin{gather*}
S_1 = 
\boxed
{
\begin{matrix}
0 & 0 & 0\\
0 & 1 & 0\\
0 & 0 & 0
\end{matrix}
}
\boxed
{
\begin{matrix}
0 & 1 & 1\\
1 & 1 & 1\\
1 & 1 & 1
\end{matrix}
}
\boxed
{
\begin{matrix}
0\\
0\\
0
\end{matrix}
}
\boxed
{
\begin{matrix}
1\\
1\\
1
\end{matrix}
},
\quad
S_2 = 
\boxed
{
\begin{matrix}
0 & 0 & 0\\
0 & 1 & 1\\
0 & 1 & 1
\end{matrix}
}
\boxed
{
\begin{matrix}
0 & 1 & 1\\
1 & 1 & 1\\
1 & 1 & 1
\end{matrix}
}
\boxed
{
\begin{matrix}
0\\
0\\
1
\end{matrix}
}
\boxed
{
\begin{matrix}
0\\
0\\
0
\end{matrix}
}
\boxed
{
\begin{matrix}
1\\
1\\
1
\end{matrix}
},\\
S_3 = 
\boxed
{
\begin{matrix}
0 & 0 & 2\\
0 & 1 & 2\\
2 & 2 & 2
\end{matrix}
}
\boxed
{
\begin{matrix}
0 & 1 & 2\\
1 & 1 & 2\\
2 & 2 & 2
\end{matrix}
}
\boxed
{
\begin{matrix}
0\\
0\\
1
\end{matrix}
}
\boxed
{
\begin{matrix}
0\\
0\\
0
\end{matrix}
}
\boxed
{
\begin{matrix}
1\\
1\\
1
\end{matrix}
},
\quad
S_4 = 
\boxed
{
\begin{matrix}
0 & 0 & 2\\
0 & 1 & 2\\
2 & 2 & 2
\end{matrix}
}
\boxed
{
\begin{matrix}
0 & 1 & 2\\
1 & 1 & 2\\
2 & 2 & 2
\end{matrix}
}
\boxed
{
\begin{matrix}
0 & 0 & 0\\
0 & 0 & 2\\
0 & 0 & 2
\end{matrix}
}
\boxed
{
\begin{matrix}
0\\
0\\
0
\end{matrix}
}
\boxed
{
\begin{matrix}
1\\
1\\
1
\end{matrix}
},\\
S_5 = 
\boxed
{
\begin{matrix}
0 & 0 & 2\\
0 & 1 & 2\\
2 & 2 & 2
\end{matrix}
}
\boxed
{
\begin{matrix}
0 & 1 & 2\\
1 & 1 & 2\\
2 & 2 & 2
\end{matrix}
}
\boxed
{
\begin{matrix}
2\\
2\\
0
\end{matrix}
}
\boxed
{
\begin{matrix}
0\\
0\\
0
\end{matrix}
}
\boxed
{
\begin{matrix}
1\\
1\\
1
\end{matrix}
}
\boxed
{
\begin{matrix}
2\\
2\\
2
\end{matrix}
},
\quad
S_6 = 
\boxed
{
\begin{matrix}
0 & 0 & 2\\
0 & 1 & 2\\
0 & 0 & 2
\end{matrix}
}
\boxed
{
\begin{matrix}
0 & 1 & 2\\
1 & 1 & 2\\
0 & 0 & 2
\end{matrix}
}
\boxed
{
\begin{matrix}
2\\
2\\
0
\end{matrix}
}.
\end{gather*}

В настоящей работе доказан критерий неявной полноты в $P_3$
в терминах предполных классов: 
система $\mathfrak{A}\subseteq P_3$ 
неявно полна тогда и только тогда,
когда она целиком не содержится ни в одном из 
54 (описанных в работе) неявно предполных классов.

\begin{statement}\label{submaxHasX}
Любой неявно предполный класс $\mathfrak{A}\subset P_k$ 
совпадает со своим явным замыканием.
\end{statement}
\begin{proof}
Предположим, что $E(\mathfrak{A})\neq\mathfrak{A}$
и найдется функция ${f\in E(\mathfrak{A}) \setminus \mathfrak{A}}$.
Тогда из определения неявно предполного класса получаем, 
что $I(\mathfrak{A}\cup\{f\}) = P_k$.
С другой стороны, 
из свойств неявной и явной выразимости следует, 
что 
$$
I(\mathfrak{A}) = I(E(\mathfrak{A})) =
I(E(\mathfrak{A}\cup \{f\})) = I(\mathfrak{A}\cup \{f\}).
$$
Но в таком случае класс $\mathfrak{A}$ оказывается неявно полным. 
Противоречие.
\end{proof}

\begin{follow}
Любой неявно предполный класс в $P_k$ 
содержит тождественную функцию.
\end{follow}

Говорят, 
что функция $k$-значной логики $f(x_1,\dots, x_n)$ 
\textit{сохраняет предикат} 
$\sigma(x_1,\dots,x_m)$, определенный на $E_k^m$, 
если для любых наборов $\tilde{\alpha}^1,\dots,\tilde{\alpha}^n$, 
где $\tilde{\alpha}^i = (\alpha^i_1,\alpha^i_2,\dots,\alpha^i_m)$, 
на которых предикат $\sigma$ принимает истинное значение,
на наборе 
$$
(f(\alpha^1_1,\alpha^2_1,\dots,\alpha^n_1),
f(\alpha^1_2,\alpha^2_2,\dots,\alpha^n_2),\dots,
f(\alpha^1_m,\alpha^2_m,\dots,\alpha^n_m))
$$
он также истинен.
Говорят, что класс функций $k$-значной логики является 
\textit{классом сохранения предиката} $\sigma$,
если он содержит те и только те функции, 
которые сохраняют предикат $\sigma$.
Класс сохранения предиката $\sigma$ 
будем обозначать через $Pol(\sigma)$.
Если класс функций можно задать как 
класс сохранения некоторого предиката,
то он называется \textit{предикатно описуемым}.

Если записать матрицу, 
столбцами которой будут являться все наборы предиката,
на которых он равен 1,
то можно определить понятие сохранения матрицы в следующем смысле.
Рассмотрим матрицу 
$A = \{a_i^j\}$, где $1\leq i\leq t, 1\leq j\leq s$, 
а элементы матрицы $a_i^j$ принадлежат множеству $E_k$.
Говорят, 
что функция $f(x_1,\dots,x_n)\in P_k$ 
\textit{сохраняет матрицу} $A$,
если для любых, быть может повторяющихся, столбцов этой матрицы
$$
\begin{pmatrix}
a_1^{j_1}\\
a_2^{j_1}\\
\dots \\
a_t^{j_1}\\
\end{pmatrix},
\begin{pmatrix}
a_1^{j_2}\\
a_2^{j_2}\\
\dots \\
a_t^{j_2}\\
\end{pmatrix},
\dots,
\begin{pmatrix}
a_1^{j_n}\\
a_2^{j_n}\\
\dots \\
a_t^{j_n}\\
\end{pmatrix}
$$
столбец 
$$
\begin{pmatrix}
f(a_1^{j_1}, a_1^{j_2}, \dots, a_1^{j_n})\\
f(a_2^{j_1}, a_2^{j_2}, \dots, a_2^{j_n})\\
\dots \\
f(a_t^{j_1}, a_t^{j_2}, \dots, a_t^{j_n})\\
\end{pmatrix}
$$
также является столбцом матрицы $A$.
Множество всех функций, 
сохраняющих матрицу $A$ называется 
\textit{классом сохранения матрицы} $A$ 
и обозначается через $Pol(A)$.

Отметим некоторые известные свойства 
классов сохранения матриц.
\begin{statement}\label{matrix}
Пусть $A$~--- матрица, элементы которой принадлежат множеству $E_k$.
Тогда

1) $Pol(A)$ является замкнутым по суперпозиции классом в $P_k$.

2) Все селекторные функции принадлежат $Pol(A)$.

3) Если матрица $A'$ получена из матрицы $A$ перестановкой строк 
и/или столбцов, 
то $Pol(A') = Pol(A)$.

4) Если матрица $A'$ получена из матрицы $A$ удалением 
одной или нескольких строк,
то $Pol(A) \subseteq Pol(A')$.
\end{statement}

В работе~\cite{Kuznetsov} доказано, 
что если класс сохраняет (функциональный) предикат вида 
$x_n = f(x_1,\dots,x_n)$, 
то класс является параметрически неполным.

\textit{1-основанием} (или просто \textit{основанием})
замкнутого по суперпозиции 
класса функций $k$-значной логики
называется множество всех одноместных функций, 
содержащихся в этом классе.
При изучении неявно предполных классов в $P_3$
основания этих классов играют важную роль.
Пусть $F$~--- замкнутый по суперпозиции класс 
одноместных функций в $P_k$. 
\textit{Максимальным надклассом} $F$ 
будем называть максимальный (по включению) 
замкнутый по суперпозиции класс,
основание которого совпадает с $F$.

Несложно доказывается следующая 
\begin{lemma}\label{pol_base}
Пусть $F=\{f_1(x),\dots,f_n(x)\}\subset P_3$~--- 
замкнутый по суперпозиции класс одноместных функций, 
содержащий тождественную функцию. 
Тогда максимальный надкласс $F$
совпадает с классом сохранения матрицы $A$, 
состоящей из столбцов вида
$
\begin{pmatrix}
f_i(0)\\
f_i(1)\\
f_i(2)
\end{pmatrix}
$ для всех $i=1\dots n$.
\end{lemma}

Доказательства следующих утверждений можно найти, 
например, в~\cite{Yablonskiy}.
\begin{statement}\label{NotIn}
1) Из функции, не сохраняющей подмножество $E\subset E_k$,
с помощью подстановки констант из множества $E$ 
и отождествления переменных
можно получить константу из множества $E_k \setminus E$.

2) Из функции,
не сохраняющей некоторое разбиение $B$,
с помощью подстановки констант $0,\dots,k-1$ и отождествления переменных
можно получить одноместную функцию,
не сохраняющую разбиение $B$.

3) Из функции,
не являющейся монотонной относительно некоторого порядка $\leq$,
с помощью подстановки констант $0,\dots,k-1$ и отождествления переменных
можно получить одноместную функцию,
не являющуюся монотонной относительно порядка $\leq$.

4) Из нелинейной функции
с помощью подстановки констант $0,\dots,k-1$ и отождествления переменных
можно получить одноместную нелинейную функцию.
\end{statement}

\begin{statement}\label{nonlin}
Функция одного переменного $f(x)$, выпускающая ровно одно 
значение, и функция $x+\alpha \pmod{3} \;(0<\alpha\le 2)$ 
порождают все функции 
одного переменного, выпускающие хоть одно значение.
\end{statement}

\section{Неявно предполные классы}
\addtocounter{subsection}{1}
Перейдем к описанию неявно предполных классов.
Обозначения предполных по суперпозиции классов в $P_3$ даны
по~\cite{Yablonskiy}.

\begin{theorem}
Классы самодвойственных функций $S$, линейных функций $L$,
а также классы $T_{\{a\},0}$ сохранения констант
являются неявно предполными.
\end{theorem}

\begin{proof}
С одной стороны, 
все эти классы являются классами сохранения функциональных предикатов
$x = y+1;\; x = y + z - t$ и $x = a$ соответственно 
(см., например,~\cite{Lau}) и, следовательно,
параметрически и неявно не полны.
С другой, при добавлении к ним любой функции не из класса
они составляют полную по суперпозиции и, следовательно,
неявно полную систему. 
\end{proof}

Следуя~\cite{Orehova_Pk},
назовем четверку различных наборов из $E_3^n$ \textit{квадратом},
если у всех наборов все компоненты кроме двух совпадают, 
а каждая из оставшихся компонент принимает ровно два значения, 
каждое из которых принимается на двух наборах.

Будем говорить, что функция $f$ 
\textit{выделяет вершину $\tilde{\alpha_i}$ квадрата}
$\tilde{\alpha}_1,\tilde{\alpha}_2,
\tilde{\alpha}_3,\tilde{\alpha}_4$,
если значение функции на $\tilde{\alpha}_i$ 
отлично от значений на остальных наборах.

Через $\mathfrak{N}$ обозначим класс всех функций в $P_3$, 
не выделяющих вершины ни одного квадрата.

\begin{theorem}\label{N_Submax}
\cite{Orehova_Pk}
Класс квазилинейных функций неявно предполон в $P_3$.
\end{theorem}

Для произвольного отношения порядка $\leq$ определим 
трехместные предикаты $\rho$ и $\rho'$:
\begin{gather*}
\rho(a,b,c)=1 \Leftrightarrow ((a \leq b) \wedge (c = b)) \vee
((b \leq a) \wedge (c = a)),
\\
\rho'(a,b,c)=1 \Leftrightarrow ((b \leq a) \wedge (c = b)) \vee
((a \leq b) \wedge (c = a)).
\end{gather*}
В работе~\cite{Starostin_M} доказано, 
что если существуют хотя бы два элемента 
сравнимых относительно порядка $\leq$, 
то классы $Pol(\rho)$ и $Pol(\rho')$ являются неявно предполными.

Для линейного порядка $0<1<2$ введем обозначения 
$DM_1 = Pol(\rho)$, $KM_1 = Pol(\rho')$.
Для порядка, в котором 2 больше, чем 0 и 1, 
которые не сравнимы между собой, обозначим 
$R_2' = Pol(\rho), Q_2' = Pol(\rho')$.
Кроме того, определим матрицы, 
соответствующие предикатам $\rho$ и $\rho'$ для такого порядка.

$$
A_1 = \begin{pmatrix}
0 & 1 & 0 & 1 & 2 & 2 & 2\\
0 & 1 & 2 & 2 & 0 & 1 & 2\\
0 & 1 & 2 & 2 & 2 & 2 & 2
\end{pmatrix}\;\quad
A_2 = \begin{pmatrix}
0 & 1 & 0 & 1 & 2 & 2 & 2\\
0 & 1 & 2 & 2 & 0 & 1 & 2\\
0 & 1 & 0 & 1 & 0 & 1 & 2
\end{pmatrix}\;\quad
$$

Следующая теорема является частным случаем результата
из~\cite{Starostin_M}
\begin{theorem}
Классы $KM_1, DM_1, R_2', Q_2'$, а также двойственные им классы
являются неявно предполными в $P_3$.
\end{theorem}

Через $F'_2$ обозначим класс сохранения матрицы 
$
A_3 =
\begin{pmatrix}
0 & 1 & 0 & 0 & 1 & 1 & 2\\
0 & 1 & 0 & 1 & 0 & 1 & 2\\
0 & 1 & 2 & 2 & 2 & 2 & 2
\end{pmatrix}
$.

\begin{theorem}\label{F_submax}
Класс функций $F'_2$, а также двойственные ему классы 
неявно предполны в $P_3$.
\end{theorem}

\begin{proof}
Выпишем все одноместные функции класса $F'_2$.

$$
\boxed
{
\begin{matrix}
0\\
0\\
0
\end{matrix}
}
\boxed
{
\begin{matrix}
0\\
0\\
2
\end{matrix}
}
\boxed
{
\begin{matrix}
0\\
1\\
2
\end{matrix}
}
\boxed
{
\begin{matrix}
1\\
0\\
2
\end{matrix}
}
\boxed
{
\begin{matrix}
1\\
1\\
1
\end{matrix}
}
\boxed
{
\begin{matrix}
1\\
1\\
2
\end{matrix}
}
\boxed
{
\begin{matrix}
2\\
2\\
2
\end{matrix}
}
$$

Покажем, что класс $F'_2$ неявно неполон.
В частности, над ним неявно невыразима функция 
$
\boxed
{
\begin{matrix}
2\\
0\\
2
\end{matrix}
}
$. Действительно, в любом ее неявном представлении должно найтись уравнение
$A(x,y)=B(x,y)$, где $A,B\in F'_2$, такое, что 
$
A(x,y)=
\boxed
{
\begin{matrix}
* & * & a  \\
b & * & h_1\\
* & * & c
\end{matrix}
}$,
$B(x,y)=
\boxed
{
\begin{matrix}
* & * & a  \\
b & * & h_2\\
* & * & c
\end{matrix}
}
$, где $h_1\neq h_2$. 
Тогда, с одной стороны, в $F'_2$ должна найтись пара функций 
$
A(1,y)=
\boxed
{
\begin{matrix}
b\\
*\\
h_1
\end{matrix}
}
,
B(1,y)=
\boxed
{
\begin{matrix}
b\\
*\\
h_2
\end{matrix}
}
$, совпадающих при $y=0$ и различающихся при $y=2$, а, следовательно, либо $h_1=2$, либо $h_2=2$.

С другой, должна быть пара функций 
$
A(x,2)=
\boxed
{
\begin{matrix}
a  \\
h_1\\
c
\end{matrix}
}
,
B(x,2)=
\boxed
{
\begin{matrix}
a  \\
h_2\\
c
\end{matrix}
}
$, различающихся только при $x=1$, а, следовательно, $h_1\neq 2$ и $h_2\neq 2$. 
Противоречие.

Докажем, что при добавлении к $F'_2$ любой функции $f\in P_3\setminus F'_2$ 
получается неявно полная система.
Пусть $f(x_1,\dots,x_n)$ не принадлежит $F'_2$ и, значит, не сохраняет 
матрицу $A_3$. 
Тогда при подстановке в $f$ столбцов из $A_3$ в 
некотором порядке результирующий столбец 
$
\begin{pmatrix}
a\\
b\\
c
\end{pmatrix}
$
не принадлежит матрице. 
Из функции $f$ посредством суперпозиции 
(отождествив переменные, соответствующие одинаковым столбцам, возможно, 
переставив некоторые переменные и добавив фиктивные) можно получить функцию
$f'(x_1,\dots,x_7)\notin F'_2$ от семи переменных 
(по числу столбцов в матрице $A_3$) такую, что 
\begin{gather*}
f'(0,1,0,0,1,1,2)=a;\\
f'(0,1,0,1,0,1,2)=b;\\
f'(0,1,2,2,2,2,2)=c.
\end{gather*}
Тогда функция
$
g(x)=
f'(0,1,
\boxed
{
\begin{matrix}
0\\
0\\
2
\end{matrix}
}(x),
x,
\boxed
{
\begin{matrix}
1\\
0\\
2
\end{matrix}
}(x),
\boxed
{
\begin{matrix}
1\\
1\\
2
\end{matrix}
}(x),
2)
$
будет иметь вид 
$
\boxed
{
\begin{matrix}
a\\
b\\
c
\end{matrix}
}
$, причем столбец
$
\begin{pmatrix}
a\\
b\\
c
\end{pmatrix}
$ не принадлежит матрице $A_3$.
По построению $g(x)\in [F'_2 \cup \{f\}]$.

Докажем, что явное замыкание системы $F'_2\cup \{g\}$ содержит в себе
неявно полную систему. 
Выпишем все столбцы
$
\begin{pmatrix}
a\\
b\\
c
\end{pmatrix}
$, не принадлежащие матрице $A_3$. 

$$
\boxed{\begin{matrix}0\\0\\1\end{matrix}}
\boxed{\begin{matrix}0\\1\\0\end{matrix}}
\boxed{\begin{matrix}0\\1\\1\end{matrix}}
\boxed{\begin{matrix}0\\2\\0\end{matrix}}
\boxed{\begin{matrix}0\\2\\1\end{matrix}}
\boxed{\begin{matrix}0\\2\\2\end{matrix}}
\boxed{\begin{matrix}1\\0\\0\end{matrix}}
\boxed{\begin{matrix}1\\0\\1\end{matrix}}
\boxed{\begin{matrix}1\\1\\0\end{matrix}}
\boxed{\begin{matrix}1\\2\\0\end{matrix}}
\boxed{\begin{matrix}1\\2\\1\end{matrix}}
\boxed{\begin{matrix}1\\2\\2\end{matrix}}
\boxed{\begin{matrix}2\\0\\0\end{matrix}}
\boxed{\begin{matrix}2\\0\\1\end{matrix}}
\boxed{\begin{matrix}2\\0\\2\end{matrix}}
\boxed{\begin{matrix}2\\1\\0\end{matrix}}
\boxed{\begin{matrix}2\\1\\1\end{matrix}}
\boxed{\begin{matrix}2\\1\\2\end{matrix}}
\boxed{\begin{matrix}2\\2\\0\end{matrix}}
\boxed{\begin{matrix}2\\2\\1\end{matrix}}
$$
Рассмотрим несколько случаев.

а) 
$
g(x)=
\boxed
{
\begin{matrix}
2\\
2\\
0
\end{matrix}
}
$.

Так как система
$S_5=
\boxed
{
\begin{matrix}
0 & 0 & 2\\
0 & 1 & 2\\
2 & 2 & 2
\end{matrix}
}
\boxed
{
\begin{matrix}
0 & 1 & 2\\
1 & 1 & 2\\
2 & 2 & 2
\end{matrix}
}
\boxed
{
\begin{matrix}
0\\
0\\
0
\end{matrix}
}
\boxed
{
\begin{matrix}
1\\
1\\
1
\end{matrix}
}
\boxed
{
\begin{matrix}
2\\
2\\
0
\end{matrix}
}
$ неявно полна, 
а первые четыре функции лежат в классе $F'_2$, то вместе с $g(x)$
получается неявно полная система.

б) 
$
g(x)\in
\left\lbrace
\boxed
{
\begin{matrix}
0\\
2\\
0
\end{matrix}
}
\boxed
{
\begin{matrix}
2\\
0\\
0
\end{matrix}
}
\boxed
{
\begin{matrix}
2\\
2\\
1
\end{matrix}
}
\right\rbrace
$

Из всех этих функций можно получить функцию
$
\boxed
{
\begin{matrix}
2\\
2\\
0
\end{matrix}
}
$, 
и этот случай сведется к предыдущему.

$$
\boxed
{
\begin{matrix}
0\\
2\\
0
\end{matrix}
}
\left(
\boxed
{
\begin{matrix}
1\\
1\\
2
\end{matrix}
}
\right)
=
\boxed
{
\begin{matrix}
2\\
2\\
0
\end{matrix}
};
\boxed
{
\begin{matrix}
2\\
0\\
0
\end{matrix}
}
\left(
\boxed
{
\begin{matrix}
0\\
0\\
2
\end{matrix}
}
\right)
=
\boxed
{
\begin{matrix}
2\\
2\\
0
\end{matrix}
};
\boxed
{
\begin{matrix}
1\\
0\\
2
\end{matrix}
}
\left(
\boxed
{
\begin{matrix}
2\\
2\\
1
\end{matrix}
}
\right)
=
\boxed
{
\begin{matrix}
2\\
2\\
0
\end{matrix}
}. 
$$

в)
$
g(x)\in 
\left\lbrace
\boxed
{
\begin{matrix}
0\\
2\\
1
\end{matrix}
}
\boxed
{
\begin{matrix}
1\\
2\\
0
\end{matrix}
}
\boxed
{
\begin{matrix}
1\\
2\\
1
\end{matrix}
}
\boxed
{
\begin{matrix}
2\\
0\\
1
\end{matrix}
}
\boxed
{
\begin{matrix}
2\\
1\\
0
\end{matrix}
}
\boxed
{
\begin{matrix}
2\\
1\\
1
\end{matrix}
}
\right\rbrace
$

$
\boxed
{
\begin{matrix}
0\\
0\\
2
\end{matrix}
}
(g(x))
$ есть либо функция 
$
\boxed
{
\begin{matrix}
0\\
2\\
0
\end{matrix}
}
$, либо функция 
$
\boxed
{
\begin{matrix}
2\\
0\\
0
\end{matrix}
}
$, и этот случай сводится к пункту б).

г) 
$
g(x)=
\boxed
{
\begin{matrix}
0\\
2\\
2
\end{matrix}
}
$.

Так как система
$
\boxed
{
\begin{matrix}
0 & 2 & 2\\
2 & 2 & 2\\
2 & 2 & 2
\end{matrix}
}
\boxed
{
\begin{matrix}
0 & 0 & 0\\
0 & 0 & 0\\
0 & 0 & 2
\end{matrix}
}
\boxed
{
\begin{matrix}
0\\
0\\
0
\end{matrix}
}
\boxed
{
\begin{matrix}
2\\
2\\
2
\end{matrix}
}
$, двойственна неявно полной системе $S_1$ относительно подстановки $(12)$, 
последние три функции лежат в классе $F'_2$, а первая выражается над
$F'_2\cup \{g\}$
$$
\boxed
{
\begin{matrix}
0 & 0 & 2\\
0 & 0 & 2\\
2 & 2 & 2
\end{matrix}
}
\left(
\boxed
{
\begin{matrix}
0\\
2\\
2
\end{matrix}
}(x),
\boxed
{
\begin{matrix}
0\\
2\\
2
\end{matrix}
}(y)
\right)
=
\boxed
{
\begin{matrix}
0 & 2 & 2\\
2 & 2 & 2\\
2 & 2 & 2
\end{matrix}
},
\quad
\boxed
{
\begin{matrix}
0 & 0 & 2\\
0 & 0 & 2\\
2 & 2 & 2
\end{matrix}
}\in F'_2
$$ 
то система $F'_2\cup \{g(x)\}$ неявно полна.

д)
$
g(x)\in
\left\lbrace
\boxed
{
\begin{matrix}
1\\
2\\
2
\end{matrix}
}
\boxed
{
\begin{matrix}
2\\
0\\
2
\end{matrix}
}
\boxed
{
\begin{matrix}
2\\
1\\
2
\end{matrix}
}
\right\rbrace
$.

Из всех этих функций можно получить функцию
$
\boxed
{
\begin{matrix}
0\\
2\\
2
\end{matrix}
}
$, и этот случай сведется к предыдущему:

$$
\boxed
{
\begin{matrix}
1\\
0\\
2
\end{matrix}
}
\left(
\boxed
{
\begin{matrix}
1\\
2\\
2
\end{matrix}
}
\right)
=
\boxed
{
\begin{matrix}
0\\
2\\
2
\end{matrix}
};
\boxed
{
\begin{matrix}
2\\
0\\
2
\end{matrix}
}
\left(
\boxed
{
\begin{matrix}
1\\
0\\
2
\end{matrix}
}
\right)
=
\boxed
{
\begin{matrix}
0\\
2\\
2
\end{matrix}
};
\boxed
{
\begin{matrix}
1\\
0\\
2
\end{matrix}
}
\left(
\boxed
{
\begin{matrix}
2\\
1\\
2
\end{matrix}
}
\right)
=
\boxed
{
\begin{matrix}
2\\
0\\
2
\end{matrix}
}.
$$

е)
$
g(x)=
\boxed
{
\begin{matrix}
0\\
0\\
1
\end{matrix}
}
$.

Так как система
$S_3=
\boxed
{
\begin{matrix}
0 & 0 & 2\\
0 & 1 & 2\\
2 & 2 & 2
\end{matrix}
}
\boxed
{
\begin{matrix}
0 & 1 & 2\\
1 & 1 & 2\\
2 & 2 & 2
\end{matrix}
}
\boxed
{
\begin{matrix}
1\\
1\\
1
\end{matrix}
}
\boxed
{
\begin{matrix}
0\\
0\\
0
\end{matrix}
}
\boxed
{
\begin{matrix}
0\\
0\\
1
\end{matrix}
}
$ неявно полна, а
первые четыре функции лежат в классе $F'_2$, то вместе с $g(x)$
получается неявно полная система.

ж)
$
g(x)\in
\left\lbrace
\boxed
{
\begin{matrix}
0\\
1\\
0
\end{matrix}
}
\boxed
{
\begin{matrix}
0\\
1\\
1
\end{matrix}
}
\boxed
{
\begin{matrix}
1\\
1\\
0
\end{matrix}
}
\right\rbrace
$.

Из всех этих функций можно получить функцию
$
\boxed
{
\begin{matrix}
0\\
0\\
1
\end{matrix}
}
$, и этот случай сведется к предыдущему.

$$
\boxed
{
\begin{matrix}
1\\
0\\
2
\end{matrix}
}
\left(
\boxed
{
\begin{matrix}
1\\
1\\
0
\end{matrix}
}
\right)
=
\boxed
{
\begin{matrix}
0\\
0\\
1
\end{matrix}
};
\boxed
{
\begin{matrix}
0\\
1\\
0
\end{matrix}
}
\left(
\boxed
{
\begin{matrix}
1\\
1\\
2
\end{matrix}
}
\right)
=
\boxed
{
\begin{matrix}
1\\
1\\
0
\end{matrix}
};
\boxed
{
\begin{matrix}
0\\
1\\
1
\end{matrix}
}
\left(
\boxed
{
\begin{matrix}
0\\
0\\
2
\end{matrix}
}
\right)
=
\boxed
{
\begin{matrix}
0\\
0\\
1
\end{matrix}
}.
$$

з) 
$
g(x)\in
\left\lbrace
\boxed
{
\begin{matrix}
1\\
0\\
0
\end{matrix}
}
\boxed
{
\begin{matrix}
1\\
0\\
1
\end{matrix}
}
\right\rbrace
$.

Из обеих функций можно получить функции из пункта ж).
$$
\boxed
{
\begin{matrix}
1\\
0\\
2
\end{matrix}
}
\left(
\boxed
{
\begin{matrix}
1\\
0\\
0
\end{matrix}
}
\right)
=
\boxed
{
\begin{matrix}
0\\
1\\
1
\end{matrix}
};
\boxed
{
\begin{matrix}
1\\
0\\
2
\end{matrix}
}
\left(
\boxed
{
\begin{matrix}
1\\
0\\
1
\end{matrix}
}
\right)
=
\boxed
{
\begin{matrix}
0\\
1\\
0
\end{matrix}
}.
$$

Таким образом, система $F'_2\cup \{g\}$ неявно полна,
следовательно, класс
$F'_2$ неявно предполон.
\end{proof}

Через $W_a$ будем обозначать пересечение классов 
$U_{\{a,b\}\{c\}}$ и $U_{\{a,c\}\{b\}}$.

\begin{theorem}\label{W_submax}
Класс $W_1$, а также двойственные ему классы 
неявно предполоны в $P_3$.
\end{theorem}

\begin{proof}
Выпишем все одноместные функции класса $W_1$
$$
\boxed
{
\begin{matrix}
0\\
0\\
0
\end{matrix}
}
\boxed
{
\begin{matrix}
0\\
1\\
1
\end{matrix}
}
\boxed
{
\begin{matrix}
0\\
1\\
2
\end{matrix}
}
\boxed
{
\begin{matrix}
1\\
0\\
0
\end{matrix}
}
\boxed
{
\begin{matrix}
1\\
1\\
1
\end{matrix}
}
\boxed
{
\begin{matrix}
1\\
1\\
2
\end{matrix}
}
\boxed
{
\begin{matrix}
2\\
2\\
1
\end{matrix}
}
\boxed
{
\begin{matrix}
2\\
2\\
2
\end{matrix}
}
$$

Покажем, что класс $W_1$ неявно неполон.
В частности, над ним неявно невыразима функция 
$
\boxed
{
\begin{matrix}
0\\
0\\
2
\end{matrix}
}
$.
Действительно, в любом ее неявном представлении 
должна найтись пара функций $g_1(x,y), g_2(x,y)$ вида
$
\boxed
{
\begin{matrix}
a & *   & *\\
b & h_1 & *\\
* & *   & c
\end{matrix}
}
$
и
$
\boxed
{
\begin{matrix}
a & *   & *\\
b & h_2 & *\\
* & *   & c
\end{matrix}
}
$ соответственно, где $h_1\neq h_2$. 
Отождествим переменные этих функций, получим, что в $W_1$ 
должна найтись пара одноместных функций, различающихся только в единице.
Из списка одноместных функций видно, что таких функций в классе нет.

Докажем, что при добавлении к $W_1$ любой функции $f\in P_3\setminus W_1$ 
получается неявно полный класс.
Пусть $f(x_1,\dots,x_n)\notin W_1$. 
Без ограничения общности будем считать, 
что $f$ не сохраняет разбиение $\{12\}\{0\}$. 
Так как все три константы лежат в 
классе $W_1$, то по утверждению~\ref{NotIn} с помощью подстановки констант 
из $f(x_1,\dots,x_n)\notin W_1$ можно получить одноместную функцию 
$g(x)\notin W_1$. 
Покажем, что явное замыкание системы $W_1\cup \{g(x)\}$ содержит в себе
одну из двух неявно полных систем:
$
S_1=
\boxed
{
\begin{matrix}
0 & 1 & 1\\
1 & 1 & 1\\
1 & 1 & 1
\end{matrix}
}
\boxed
{
\begin{matrix}
0 & 0 & 0\\
0 & 1 & 0\\
0 & 0 & 0
\end{matrix}
}
\boxed
{
\begin{matrix}
0\\
0\\
0
\end{matrix}
}
\boxed
{
\begin{matrix}
1\\
1\\
1
\end{matrix}
}
$ или
$S_2=
\boxed
{
\begin{matrix}
0 & 1 & 1\\
1 & 1 & 1\\
1 & 1 & 1
\end{matrix}
}
\boxed
{
\begin{matrix}
0 & 0 & 0\\
0 & 1 & 1\\
0 & 1 & 1
\end{matrix}
}
\boxed
{
\begin{matrix}
0\\
0\\
1
\end{matrix}
}
\boxed
{
\begin{matrix}
1\\
1\\
1
\end{matrix}
}
$.

Так как функции 
$
\boxed
{
\begin{matrix}
0 & 1 & 1\\
1 & 1 & 1\\
1 & 1 & 1
\end{matrix}
}
$ и
$
\boxed
{
\begin{matrix}
0 & 0 & 0\\
0 & 1 & 1\\
0 & 1 & 1
\end{matrix}
}
$ принадлежат классу $W_1$, а 
$$
\boxed
{
\begin{matrix}
0 & 0 & 0\\
0 & 1 & 0\\
0 & 0 & 0
\end{matrix}
}(x,y) =
\boxed
{
\begin{matrix}
0 & 0 & 0\\
0 & 1 & 1\\
0 & 1 & 1
\end{matrix}
}
\left(
\boxed
{
\begin{matrix}
0\\
1\\
0
\end{matrix}
}(x),
\boxed
{
\begin{matrix}
0\\
1\\
0
\end{matrix}
}(y)
\right),
$$
то для неявной полноты нам достаточно получить одну из функций
$
\boxed
{
\begin{matrix}
0\\
0\\
1
\end{matrix}
}
\boxed
{
\begin{matrix}
0\\
1\\
0
\end{matrix}
}
$.

Далее покажем, что по крайней мере одну из этих функций можно выразить через
$g(x)$ и функции 
$
\boxed
{
\begin{matrix}
0\\
1\\
1
\end{matrix}
}
\boxed
{
\begin{matrix}
1\\
0\\
0
\end{matrix}
}
$, лежащие в классе $W_1$.
Так как $g(x)$ не сохраняет разбиение $(12)(0)$, то она имеет вид
$
\boxed
{
\begin{matrix}
*\\
0\\
a
\end{matrix}
}
$ или 
$
\boxed
{
\begin{matrix}
*\\
b\\
0
\end{matrix}
}
$, $a,b\in\{1,2\}$.
Пусть 
$
g(x)=
\boxed
{
\begin{matrix}
*\\
0\\
a
\end{matrix}
}
$. Если $*$ --- это $0$, то
$
\boxed
{
\begin{matrix}
0\\
1\\
1
\end{matrix}
}(g(x))=
\boxed
{
\begin{matrix}
0\\
0\\
1
\end{matrix}
}
$. 
Если же $*$ --- это $1$ или $2$, то
$
\boxed
{
\begin{matrix}
1\\
0\\
0
\end{matrix}
}(g(x))=
\boxed
{
\begin{matrix}
0\\
1\\
0
\end{matrix}
}
$.
Случай, когда $g$ имеет вид 
$
\boxed
{
\begin{matrix}
*\\
b\\
0
\end{matrix}
}
$, рассматривается аналогично.

Таким образом, система $W_1\cup \{g\}$ неявно полна,
и, следовательно, класс $W_1$ является неявно предполным.
\end{proof}

Через $Y_2$ обозначим пересечение трех 
предполных по суперпозиции классов
трехзначной логики: $T_{\{0,1\},0}, U_{\{0,1\}\{2\}}$ и
$T_{\{2\},1}$.

\begin{theorem}\label{Y_submax}
Класс $Y_2$, а также двойственные ему классы 
неявно предполны в $P_3$.
\end{theorem}
\begin{proof}
Выпишем все одноместные функции класса $Y_2$.
$$
\boxed
{
\begin{matrix}
0\\
0\\
0
\end{matrix}
}
\boxed
{
\begin{matrix}
0\\
0\\
2
\end{matrix}
}
\boxed
{
\begin{matrix}
0\\
1\\
2
\end{matrix}
}
\boxed
{
\begin{matrix}
1\\
0\\
2
\end{matrix}
}
\boxed
{
\begin{matrix}
1\\
1\\
1
\end{matrix}
}
\boxed
{
\begin{matrix}
1\\
1\\
2
\end{matrix}
}
$$

Покажем, что класс $Y_2$ неявно неполон.
В частности, над ним неявно невыразима функция 
$
\boxed
{
\begin{matrix}
0\\
1\\
0
\end{matrix}
}
$. Действительно, в любом ее неявном представлении должна найтись 
пара функций $f_1(x,y),f_2(x,y)\in Y_2$ вида 
$
\boxed
{
\begin{matrix}
a & *   & *\\
* & b   & *\\
c & h_1 & *
\end{matrix}
}$ и
$
\boxed
{
\begin{matrix}
a & *   & *\\
* & b   & *\\
c & h_2 & *
\end{matrix}
}
$ соответственно, где $h_1\neq h_2$. 
Тогда $c,h_1,h_2\in \{0,1\}$, 
так как в противном случае одна из функций не сохраняла бы 
разбиение $\{0,1\}\{2\}$. 
Без ограничения общности можно считать, 
что $c=h_1=0, h_2=1$.

По свойству функций из класса $T_{\{2\},1}$ 
каждая из этих функций на наборах 
$(1,1);$ $(1,2);$ $(2,1);$ $(2,2)$ 
должна выпускать одно из значений $0,1$. 
Это возможно только в том случае, если $b=2$, 
но тогда ни одна из этих функций не сохраняет подмножество $\{0,1\}$. 
Противоречие.

Докажем, что при добавлении к $Y_2$ любой функции $f\in P_3\setminus Y_2$ 
получается неявно полная система.
Рассмотрим три возможности для функции $f$:

а) $f\notin T_{\{0,1\},0}$

Согласно утверждению~\ref{NotIn} подстановкой констант $0$ и $1$ 
из такой функции можно получить константу $2$.
Рассмотрим функцию 
$
h(x,y)=
\boxed
{
\begin{matrix}
0 & 0 & 2\\
0 & 0 & 2\\
0 & 0 & 0
\end{matrix}
}
\in Y_2
$. Тогда
$
h(x,2)=
\boxed
{
\begin{matrix}
2\\
2\\
0
\end{matrix}
}
$.
Таким образом над $Y_2\cup \{f\}$ явно выразима неявно полная система
$S_5=
\boxed
{
\begin{matrix}
0 & 0 & 2\\
0 & 1 & 2\\
2 & 2 & 2
\end{matrix}
}
\boxed
{
\begin{matrix}
0 & 1 & 2\\
1 & 1 & 2\\
2 & 2 & 2
\end{matrix}
}
\boxed
{
\begin{matrix}
2\\
2\\
0
\end{matrix}
}
\boxed
{
\begin{matrix}
0\\
0\\
0
\end{matrix}
}
\boxed
{
\begin{matrix}
1\\
1\\
1
\end{matrix}
}
\boxed
{
\begin{matrix}
2\\
2\\
2
\end{matrix}
}
$ 
(все функции кроме третьей лежат в классе $Y_2$).

б) $f\in T_{\{0,1\},0}; f\notin U_{\{0,1\}\{2\}}$

Так как $f(\tilde{x})\notin U_{\{0,1\}\{2\}}$, существует блок,
на котором $f$ принимает и значение $2$, и значения из $\{0,1\}$.
В этом блоке найдутся два соседних набора $\tilde{\alpha}, \tilde{\beta}$
такие, что $f(\tilde{\alpha})=2, f(\tilde{\beta})\in \{0,1\}$.
Без ограничения общности можно считать, что
$$
\begin{array}{ccl}
\tilde{\alpha} = (0,0,\dots,0,a,1,1,\dots,1,2,2,\dots,2),\\
\tilde{\beta}  = (0,0,\dots,0,b,1,1,\dots,1,2,2,\dots,2),
\end{array}
$$
где $a,b\in \{0,1\}, a\neq b$.

Рассмотрим $g(x,y)=f(0,0,\dots,0,x,1,1,\dots,1,y,y,\dots,y)$. По построению
$g(a,2)=2, g(b,2)=c\in \{0,1\}$. Так как $g(x,y)\in T_{\{0,1\},0}$, то,
в зависимости от значения $a$, одна из функций
$
\boxed
{
\begin{matrix}
1\\
1\\
2
\end{matrix}
}
(g(x,y)),
\boxed
{
\begin{matrix}
1\\
1\\
2
\end{matrix}
}
(g(
\boxed
{
\begin{matrix}
1\\
0\\
2
\end{matrix}
}
x,y))
$ имеет вид
$
\boxed
{
\begin{matrix}
1 & 1 & 2\\
1 & 1 & 1\\
d_1 & d_2 & d_3
\end{matrix}
}=g'(x,y)
$, где
$
d_1,d_2,d_3\in \{1,2\}
$, причем $g'\in [Y_2\cup \{f\}]$.

Из функции $g'$ и 
$
\boxed
{
\begin{matrix}
1 & 1 & 1\\
1 & 1 & 1\\
1 & 1 & 2
\end{matrix}
}
$, принадлежащей $Y_2$, можно получить функцию
$
\boxed
{
\begin{matrix}
1 & 1 & 2\\
1 & 1 & 1\\
1 & 1 & d_3
\end{matrix}
}=
\boxed
{
\begin{matrix}
1 & 1 & 1\\
1 & 1 & 1\\
1 & 1 & 2
\end{matrix}
}(
\boxed
{
\begin{matrix}
1\\
1\\
2
\end{matrix}
}(y),
g'(x,y))
$.

Если $d_3=2$, то над множеством $Y_2\cup \{g'\}$ 
явно выразима неявно полная система
$
\boxed
{
\begin{matrix}
0 & 0 & 2\\
0 & 1 & 2\\
2 & 2 & 2
\end{matrix}
}
\boxed
{
\begin{matrix}
0 & 1 & 2\\
1 & 1 & 2\\
2 & 2 & 2
\end{matrix}
}
\boxed
{
\begin{matrix}
1 & 1 & 2\\
1 & 1 & 1\\
1 & 1 & 2
\end{matrix}
}
\boxed
{
\begin{matrix}
0\\
0\\
0
\end{matrix}
}
\boxed
{
\begin{matrix}
1\\
1\\
1
\end{matrix}
}
$, двойственная системе $S_4$ относительно подстановки $(01)$
(все функции кроме третьей лежат в классе $Y_2$). 

Если же $d_3=1$, выразим функцию
$
\boxed
{
\begin{matrix}
1 & 1 & 2\\
1 & 1 & 1\\
1 & 1 & 2
\end{matrix}
}
$
через $g'$ и лежащие в $Y_2$
$
\boxed
{
\begin{matrix}
1 & 1 & 2\\
1 & 1 & 2\\
2 & 2 & 2
\end{matrix}
}
$ и 
$
\boxed
{
\begin{matrix}
1 & 1 & 1\\
1 & 1 & 1\\
1 & 1 & 2
\end{matrix}
}
$:

$$
\boxed
{
\begin{matrix}
1 & 1 & 2\\
1 & 1 & 2\\
2 & 2 & 2
\end{matrix}
}
\left(
\boxed
{
\begin{matrix}
1 & 1 & 1\\
1 & 1 & 1\\
1 & 1 & 2
\end{matrix}
},
\boxed
{
\begin{matrix}
1 & 1 & 2\\
1 & 1 & 1\\
1 & 1 & 1
\end{matrix}
}
\right)
=
\boxed
{
\begin{matrix}
1 & 1 & 2\\
1 & 1 & 1\\
1 & 1 & 2
\end{matrix}
}.
$$ Cнова выразили ту же неявно полную систему.

в) $f\in T_{\{0,1\},0}\cap U_{\{0,1\}\{2\}};
f\notin T_{\{2\},1}$

Так как $f\notin T_{\{2\},1}$, $f$ не сохраняет матрицу
$
\begin{pmatrix}
0 & 1 & 2 & 2 & 2 & 0 & 1\\
0 & 1 & 2 & 0 & 1 & 2 & 2
\end{pmatrix}
$, то есть при подстановке в $f$ столбцов этой матрицы в некотором порядке
результирующий столбец совпадает с
$
\begin{pmatrix}
0\\
1
\end{pmatrix}
$ или
$
\begin{pmatrix}
1\\
0
\end{pmatrix}
$.
Отождествив переменные, соответствующие одинаковым столбцам 
(и, возможно, добавив несущественные переменные), получим функцию
$f'(x_1,\dots,x_7) \notin T_{\{2\},1}$.

Рассмотрим 
$
f''(x,y,z)=f'(0,1,x,y,
\boxed
{
\begin{matrix}
1\\
1\\
2
\end{matrix}
}(y),z,
\boxed
{
\begin{matrix}
1\\
1\\
2
\end{matrix}
}(z))
$. Тогда
$f''(2,2,0)=a$, $f''(2,0,2)=b$, где $a,b\in \{0,1\}, a\neq b$.
Обозначим $f''(0,0,0)$ через $c$. Так как $f$ сохраняет подмножество $\{0,1\}$,
$c\in \{0,1\}$.

Теперь рассмотрим пару функций 
$
g(x)=f''(x,x,0)=
\boxed
{
\begin{matrix}
c\\
d_1\\
a
\end{matrix}
}
$ и 
$
h(x)=f''(x,0,x)=
\boxed
{
\begin{matrix}
c\\
d_2\\
b
\end{matrix}
}
$, где $d_1,d_2\in \{0,1\}$. 
Без ограничения общности можно считать, что $c=0$.
В противном случае можно рассмотреть функции
$
\boxed
{
\begin{matrix}
1\\
0\\
2
\end{matrix}
}(g(x))
$ и
$
\boxed
{
\begin{matrix}
1\\
0\\
2
\end{matrix}
}(h(x))
$, которые будут иметь такой же вид, и для них $c$ будет равно $0$.
Заметим, что при построении $g$ и $h$ мы использовали только функцию $f$ и
одноместные функции из класса $Y_2$. 
Следовательно, $g,h\in [Y_2\cup \{f\}]$.

Таким образом, 
для пары функций $g$ и $h$ имеется всего 4 возможности:

$$
\boxed
{
\begin{matrix}
0\\
0\\
0
\end{matrix}
}
\boxed
{
\begin{matrix}
0\\
0\\
1
\end{matrix}
};
\boxed
{
\begin{matrix}
0\\
0\\
0
\end{matrix}
}
\boxed
{
\begin{matrix}
0\\
1\\
1
\end{matrix}
};
\boxed
{
\begin{matrix}
0\\
1\\
0
\end{matrix}
}
\boxed
{
\begin{matrix}
0\\
0\\
1
\end{matrix}
};
\boxed
{
\begin{matrix}
0\\
1\\
0
\end{matrix}
}
\boxed
{
\begin{matrix}
0\\
1\\
1
\end{matrix}
}.
$$
В любом случае, 
одна из функций $g, h$ совпадает с одной из функций

$
\boxed
{
\begin{matrix}
0\\
0\\
1
\end{matrix}
}
\boxed
{
\begin{matrix}
0\\
1\\
1
\end{matrix}
}
$. Кроме того, 
$
\boxed
{
\begin{matrix}
0\\
1\\
1
\end{matrix}
}
\left(
\boxed
{
\begin{matrix}
0\\
0\\
2
\end{matrix}
}
\right)
=
\boxed
{
\begin{matrix}
0\\
0\\
1
\end{matrix}
}
$, поэтому в любом случае можно выразить функцию 
$
\boxed
{
\begin{matrix}
0\\
0\\
1
\end{matrix}
}
$.

Таким образом, 
над $Y_2\cup \{f\}$ явно выразима неявно полная система 
$$S_3=
\boxed
{
\begin{matrix}
0 & 0 & 2\\
0 & 1 & 2\\
2 & 2 & 2
\end{matrix}
}
\boxed
{
\begin{matrix}
0 & 1 & 2\\
1 & 1 & 2\\
2 & 2 & 2
\end{matrix}
}
\boxed
{
\begin{matrix}
0\\
0\\
1
\end{matrix}
}
\boxed
{
\begin{matrix}
0\\
0\\
0
\end{matrix}
}
\boxed
{
\begin{matrix}
1\\
1\\
1
\end{matrix}
},
$$ и, следовательно, класс $Y_2$ неявно предполон в $P_3$.
\end{proof}

Пусть функция $f(\tilde{x})\in P_3$ 
сохраняет подмножество $\{a,b\}$.
Рассмотрим пару отображений $\theta$ и $\theta^{-1}$ таких, что
$$
\theta \colon 
\begin{cases}
0 \to a,\\
1 \to b.
\end{cases}
\theta^{-1} \colon 
\begin{cases}
a \to 0,\\
b \to 1.
\end{cases}
$$
Опеределим булеву функцию $\hat{f}({\tilde{x}})$ следующим образом:
$$
\hat{f}(x_1,\dots,x_n)=
\theta^{-1}(f(\theta(x_1),\dots,\theta(x_n))).
$$ 
Такое определение корректно в силу того, 
что $f$ сохраняет $\{a,b\}$.
Функцию $\hat{f}$ будем называть 
\textit{булевым ограничением функции} $f$.
 
Пусть $\mathfrak{B}\subseteq P_3$ --- замкнутый класс, 
все функции которого сохраняют подмножество $\{a,b\}$. 
\textit{Булевым ограничением класса} назовем множество, 
состоящее из булевых ограничений всех функций из этого класса.

Пусть $\mathfrak{A}$ --- замкнутый класс булевых функций. 
Наибольший (по включению) подкласс функций $T_{\{a,b\},0}$ 
т.ч. его булево ограничение совпадает с $\mathfrak{A}$, 
будем обозначать $\Sigma^{\{a,b\}}_\mathfrak{A}$.

В работе~\cite{Starostin_T} доказана следующая
\begin{theorem}\label{Tsubmax}
Пусть $\mathfrak{A}$ --- один из классов 
$K,D,S,L\subset P_2$. 
Тогда классы $\Sigma^{\{0,1\}}_\mathfrak{A}$,
а также двойственные им классы являются неявно предполными в $P_3$.
\end{theorem}

Кроме того, нам понадобится доказанная в той же работе 
\begin{lemma}\label{Tbool}
Пусть $f(\tilde{x}), g(\tilde{x})\in T_{\{a,b\},0}$; 
$\hat{f}(\tilde{x})$ и $\hat{g}(\tilde{x})$ --- булевы ограничения,
соответствующие этим функциям, 
а булева функция $\hat{h}$ лежит в замыкании $\hat{f}$ и $\hat{g}$.
Тогда найдется такая функция $h(\tilde{x})\in [\{f,g\}]$, 
что ее булево ограничение есть $\hat{h}$.
\end{lemma}

На множестве наборов $E^n_3$ можно определить 
следующее отношение эквивалентности. 
Наборы $\tilde{\alpha}$ и $\tilde{\beta}$ называются 
\textit{эквивалентными относительно разбиения} $\{a,b\}\{c\}$, 
если компоненты наборов с одинаковыми номерами либо обе равны $c$, 
либо обе принадлежат множеству $\{a,b\}$. 
Класс эквивалентности относительно разбиения называется 
\textit{блоком эквивалентных наборов} или просто \textit{блоком}. 
Очевидно, что все наборы одного блока 
либо не имеют компонент равных $c$
(такой блок будем называть \textit{главным блоком}), 
либо номера этих компонент у этих наборов совпадают. 
Поэтому любой блок можно задать, 
указав номера $i_1,\dots,i_m$ тех компонент, 
которые у всех наборов отличны от $c$.
Заметим, что число наборов в блоке имеет вид $2^m$.

Рассмотрим произвольную функцию $f(x_1, \dots, x_n)$, сохраняющую разбиение 
$\{a,b\}\{c\}$~\cite{Yablonskiy}. 
Пусть на некотором блоке $B$ функция $f$ принимает 
значения из множества $\{a,b\}$. Без ограничения общности, пусть $B$ состоит
из всех наборов, отличных от $c$ на первых $m$ компонентах.
Обозначим через $f_B$ функцию
$$
f_B(x_1,\dots,x_m) = \left.f\right|_B(x_1,\dots,x_m,c,\dots,c),
$$
где $\left.f\right|_B(x_1,\dots,x_n)$ --- ограничение функции $f$ на блок $B$.
Так же, как и в определении булева 
ограничения рассмотрим функции $\theta$ и $\theta^{-1}$.
$$
\theta \colon 
\begin{cases}
0 \to a,\\
1 \to b.
\end{cases}
\theta^{-1} \colon 
\begin{cases}
a \to 0,\\
b \to 1.
\end{cases}
$$
Функцию $\hat{f}_B$ такую, что
$$
\hat{f}_B(x_1,\dots,x_n)=
\theta^{-1}(f_B(\theta(x_1),\dots,\theta(x_n))),
$$
будем называть 
\textit{булевым ограничением функции $f$ на блок $B$}.

Пусть $\mathfrak{A}$ --- 
замкнутый по суперпозиции класс булевых функций. 
Через $\Sigma^{\{a,b\}\{c\}}_\mathfrak{A}$ 
обозначим множество всех функций из $P_3$, 
сохраняющих разбиение $\{a,b\}\{c\}$, 
у которых все булевы ограничения на блоках 
принадлежат классу $\mathfrak{A}$.

\begin{note}
Константа $c$ принадлежит любому множеству вида 
$\Sigma^{\{a,b\}\{c\}}_\mathfrak{A}$.
\end{note}

\begin{lemma} \label{Umatrix}
Пусть класс $\mathfrak{A}$ булевых функций задается 
булевой матрицей 
$A$, которая содержит 
$
\left(
\begin{smallmatrix}
0 & 1\\
1 & 1
\end{smallmatrix}
\right)
$
или 
$
\left(
\begin{smallmatrix}
0 & 0\\
0 & 1
\end{smallmatrix}
\right)
$ 
в качестве своей подматрицы.
Тогда множество $\Sigma^{\{0,1\}\{2\}}_\mathfrak{A}\subset P_3$ 
совпадает с классом сохранения матрицы
$A'$,  полученной из $A$ добавлением столбца 
$
\left(
\begin{smallmatrix}
2\\
\vdots\\
2
\end{smallmatrix}
\right)
$
\end{lemma}
\begin{proof}
Докажем вложения в обе стороны.

1. Покажем, что 
$\Sigma^{\{0,1\}\{2\}}_\mathfrak{A}\subseteq Pol(A')$.

Пусть в множестве $\Sigma^{\{0,1\}\{2\}}_\mathfrak{A}$ 
найдется функция $f$, 
которая не сохраняет $A'$. 
Это значит, 
что можно подставить в $f$ столбцы из $A'$ таким образом, 
что результирующий столбец $a$ не принадлежит этой матрице. 
Рассмотрим такой набор столбцов $\tilde{a}=(a_1,\dots,a_n)$. 
Те компоненты, в которых находятся столбцы 
$
\left(
\begin{smallmatrix}
2\\
\vdots\\
2
\end{smallmatrix}
\right)
$
определяют некоторый блок наборов. 
Так как столбец $a$ отличен от столбца двоек 
(потому что $a$ не принадлежит $A'$), то корректно определено
булево ограничение $f$ на этот блок --- функция $\hat{f}$. 
С одной стороны, 
$\hat{f}\in \mathfrak{A}$, а значит сохраняет матрицу $A$. 
С другой, 
если мы подставим в $\hat{f}$ столбцы из набора $\tilde{a}$ 
(за исключением столбцов 
$
\left(
\begin{smallmatrix}
2\\
\vdots\\
2
\end{smallmatrix}
\right)
$
) в том же порядке, то в результате снова получим столбец $a$, 
который не принадлежит матрице $A$. 
Противоречие.
 
2. Покажем, что $Pol(A')\subseteq \Sigma^{\{0,1\}\{2\}}_\mathfrak{A}$.

Без ограничения общности будем считать, что $A'$ содержит подматрицу 
$
\left(
\begin{smallmatrix}
0 & 1\\
1 & 1
\end{smallmatrix}
\right)
$.
Докажем, что всякая функция $f$, сохраняющая матрицу $A'$, также
сохраняет разбиение ${\{0,1\}}\{2\}$. 
Предположим противное: пусть на некотором блоке $f$ 
принимает неэквивалентные значения. 
Без ограничения общности будем считать, что
значение на наборе
$(2, \ldots, 2, \alpha_{m+1}, \ldots, \alpha_n), 
\alpha_i\in {\{0,1\}}$ 
не эквивалентно значению на наборе $(2, \ldots, 2, 1, \ldots, 1)$. 
Подставим тогда в $f$ столбцы 
$
\left(
\begin{smallmatrix}
2\\
\vdots \\
2
\end{smallmatrix}
\right)
$
на те места, где в обоих наборах стоят двойки, столбцы, содержащие 
$
\left(
\begin{smallmatrix}
0\\
1
\end{smallmatrix}
\right)
$
на те места, где в наборе 
$\tilde{\alpha}=(\alpha_{m+1},\dots,\alpha_n)$ 
стоят нули, и
$
\left(
\begin{smallmatrix}
1\\
1
\end{smallmatrix}
\right)
$
на те места, где в наборе $\tilde{\alpha}$ стоят единицы. 
Результирующий столбец должен содержать как значение $2$, 
так и значение из набора ${\{0,1\}}$. Но таких столбцов в матрице $A'$ нет. 
Следовательно предположение неверно, и функция $f$ сохраняет разбиение.

Далее покажем, что $f\in \Sigma^{\{0,1\}\{2\}}_\mathfrak{A}$.
Рассмотрим множество булевых ограничений 
функции $f$ на всевозможные блоки. 
Все они должны сохранять матрицу $A$, 
и поэтому принадлежать классу 
$\mathfrak{A}\subseteq P_2$. 
Но тогда по определению класса $\Sigma^{\{0,1\}\{2\}}_\mathfrak{A}$
такая функция ему принадлежит.
\end{proof}

\begin{follow}
Классы вида $\Sigma^{\{0,1\}\{2\}}_\mathfrak{A}$ 
замкнуты по суперпозиции.
\end{follow}

\begin{lemma}\label{Unotfull} 
Пусть замкнутый класс $\mathfrak{A}\subseteq P_2$ неявно неполон. 
Тогда $\Sigma^{\{0,1\}\{2\}}_\mathfrak{A}$ тоже неявно неполон.
\end{lemma}

\begin{proof}
Пусть $\hat{g}$ --- некоторая булева функция, не лежащая в $I(\mathfrak{A})$,
и пусть $g\in U_{\{0,1\},\{2\}}\subseteq P_3$  --- такая функция, 
что ее булево ограничение на главном блоке совпадает с $\hat{g}$. 
Предположим, что класс $\Sigma^{\{0,1\}\{2\}}_{\mathfrak{A}}$ неявно полон. 
Тогда над $\Sigma^{\{0,1\}\{2\}}_{\mathfrak{A}}$ 
можно построить неявное представление функции $g$. 
Будем считать, что оно состоит из уравнений 
$A_i(x_1,\dots,x_n,z)=B_i(x_1,\dots,x_n,z), 1\le i\le m$.

Сперва докажем, что найдется такое число $j$, что 
у обеих функций $A_j,B_j$ определены булевы ограничения на главный блок.

Функция $g$ на наборе $(0,\dots,0)$ 
принимает некоторое булево значение $a$.
Соответственно, 
все уравнения ее неявного представления должны удовлетворяться
на наборе $(0,\dots,0,a)$.
Отсюда получаем, 
что в неявном представлении $g$ не может присутствовать уравнение,
одна часть которого имеет булево ограничение на главный блок, 
а вторая нет.
Кроме того, 
если в неявном представлении ни одна функция не имеет 
булева ограничения на главный блок, 
то все уравнения удовлетворяются 
как на наборе $(0,0,\dots,0,0)$, 
так и на наборе $(0,0,\dots,0,1)$. 
Такая система не может быть эквивалентна уравнению
$z=g(x_1,\dots,x_n)$ и,
следовательно, не является неявным представлением функции $g$.
Таким образом, 
хотя бы в одном из уравнений обе части имеют булевы 
ограничения на главный блок.

Оставим в системе только такие уравнения.
Заменим каждую функцию на ее булево ограничение на главный блок,
получим некоторую систему уравнений над $\mathfrak{A}$. 
Утверждается, что она является неявным представлением для функции 
$\hat{g}$. 

По определению любое уравнение 
из неявного представления функции $g$ 
удовлетворяется при подстановке набора
$(\tilde{\alpha},g(\tilde{\alpha}))$,
где $\tilde{\alpha}$ --- произвольный набор из $E_2^n$. 
Следовательно, 
любое уравнение полученной системы для $\hat{g}$ 
обладает этим свойством. 
Покажем, что для любого набора $\tilde{\alpha}\in E_2^n$ 
и для любого $b\in E_2$ такого, 
что $b\neq g(\tilde{\alpha})$, в новой системе найдется уравнение,
которое не выполняется при подстановке в него набора
$(\tilde{\alpha},b)$.

Действительно, 
такое уравнение должно было найтись в неявном представлении $g$.
Это не могло быть уравнение, в котором только одна из частей имеет 
булево ограничение на главный блок (таких нет).
Также это не могло быть уравнение, 
в котором ни одна из функций не имеет 
булева ограничения на главный блок, 
так как такие уравнения удовлетворяются на любом булевом наборе. 
Соответственно, 
обе функции этого уравнения имели ограничения на главный блок 
и, следовательно, в новой системе также найдется уравнение, 
которое не выполняется при подстановке в него набора
$(\tilde{\alpha},b)$.

Таким образом, полученная система 
уравнений над $\mathfrak{A}$ удовлетворяется при подстановке в нее 
наборов $(\tilde{\alpha},g(\tilde{\alpha}))$, и только их. 
Следовательно, 
эта система является неявным представлением функции $\hat{g}$, 
что противоречит предположению $\hat{g}\notin I(\mathfrak{A})$.
\end{proof}

Докажем, 
что некоторые классы вида 
$\Sigma^{\{0,1\}{\{2\}}}_\mathfrak{A}$ 
являются неявно предполными. 
Для этого нам понадобится вспомогательная 

\begin{lemma}\label{Ubool}
Пусть $f(\tilde{x}), g(\tilde{x})\in P_3$ 
сохраняют разбиение $\{0,1\}\{2\}$, 
и пусть определены 
$\hat{f}(\tilde{x}), \hat{g}(\tilde{x})\in P_2$ --- 
соответствующие булевы ограничения на главный блок, 
а булева функция 
$\hat{h}(\tilde{x})$ выразима по суперпозиции над множеством 
$\{\hat{f},\hat{g}\}$. 
Тогда в замыкании системы $\{f,g\}$ найдется 
функция $h(\tilde{x})$, 
булево ограничение которой на главный блок есть $\hat{h}$.
\end{lemma}

\begin{proof}
Так как у функций $f(\tilde{x})$ и $g(\tilde{x})$ определены 
булевы ограничения на главный блок, 
то $f$ и $g$ не принимают на нем значение $2$.
Тогда $f,g\in T_{\{0,1\},0}$. 
Воспользовавшись леммой \ref{Tbool}, 
получим искомую функцию $h(\tilde{x})\in [\{f,g\}]$.
\end{proof}

\begin{follow}
Пусть $g(\tilde{x})\in U_{\{a,b\}\{c\}}$, 
$\hat{g}(\tilde{x})$ --- булево ограничение функции $g$ на главный блок.
Пусть $\mathfrak{A}$ --- замкнутый класс булевых функций и
$[\mathfrak{A}\cup~\{\hat{g}\}]=\mathfrak{B}$.
Тогда для любой функции $\hat{h}(\tilde{x})\in \mathfrak{B}$ 
найдется такая функция 
$h(\tilde{x})\in [\Sigma^{\{a,b\}\{c\}}_\mathfrak{A}\cup \{g\}]$, 
что ее булево ограничение на главный блок есть $\hat{h}$.
\end{follow}

\begin{note}
Вообще говоря, 
эти утверждения верны не только для ограничений на главный блок.
\end{note}

\begin{theorem}\label{TKDL}
Пусть $\mathfrak{A}$ --- один из классов $K,D,L\subset P_2$. 
Тогда классы $\Sigma^{\{0,1\}\{2\}}_\mathfrak{A}$, 
а также двойственные им классы являются
неявно предполными в $P_3$.
\end{theorem}
\begin{proof}
Так как все классы $K,D,L\subset P_2$ неявно неполны, 
то по лемме \ref{Unotfull} все классы
$\Sigma^{\{0,1\}\{2\}}_\mathfrak{A}$, 
где $\mathfrak{A}\in \{K,D,L\}$ также неявно неполны.

Докажем, что при добавлении к $\Sigma^{\{0,1\}\{2\}}_\mathfrak{A}$ 
любой функции $f$ не из этого класса система становится неявно полной.

Если функция $f$ не сохраняет разбиение $\{0,1\}\{2\}$, рассмотрим 
функцию 
$\theta(x,y)=
\boxed
{
\begin{matrix}
2 & 2 & 0\\
2 & 2 & 1\\
0 & 0 & 1
\end{matrix}
}
$. Она принадлежит классу $\Sigma^{\{0,1\}\{2\}}_\mathfrak{A}$
для любого $\mathfrak{A}\in K,D,L$. 
Непосредственно проверяется, что она не лежит
ни в одном из предполных по суперпозиции классов, кроме 
$U_{\{0,1\}\{2\}}$. Но так как функция $f$ не принадлежит 
$U_{\{0,1\}\{2\}}$, $f$ и $\theta$ 
по суперпозиции порождают все $P_3$ и, следовательно, 
образуют неявно полную систему.

Если функция $f$ сохраняет разбиение, то из условия 
$f\notin \Sigma^{\{0,1\}\{2\}}_\mathfrak{A}$ следует, что
ее булево ограничение на некотором 
блоке  $B$ не лежит в классе $\mathfrak{A}\subset P_2$. 
Подставив константу $2$
(которая принадлежит всем рассматриваемым классам) 
в функцию $f$ на те места, 
где у наборов из $B$ стоит двойка, получим функцию $f'$, 
у которой булево ограничение на главном блоке не лежит в классе
$\mathfrak{A}$. 
Булево ограничение $f'$ на главном блоке обозначим через $\hat{f'}$. 

Как и в доказательстве теоремы \ref{Tsubmax}
$M\subseteq [\mathfrak{A}\cup \{\hat{f'}\}]$.
Значит, по следствию из леммы \ref{Ubool} в 
$\left[\Sigma^{\{0,1\}\{2\}}_\mathfrak{A}\cup \{f'\}\right]$
найдется функция 
с любым монотонным булевым ограничением. 
В частности, там найдется пара функций вида
$$
\&(x,y)=
\boxed
{
\begin{matrix}
0 & 0 & *\\
0 & 1 & *\\
* & * & *
\end{matrix}
}\quad
\vee(x,y)=
\boxed
{
\begin{matrix}
0 & 1 & *\\
1 & 1 & *\\
* & * & *
\end{matrix}
},
$$
из которых с помощью лежащих в $\Sigma^{\{0,1\}\{2\}}_\mathfrak{A}$ функций 
$
\boxed
{
\begin{matrix}
0\\
1\\
0
\end{matrix}
}
$ и 
$
\boxed
{
\begin{matrix}
0\\
1\\
1
\end{matrix}
}
$ нетрудно построить неявно полную систему $S_1$.
$$
\boxed
{
\begin{matrix}
0 & 0 & 0\\
0 & 1 & 0\\
0 & 0 & 0
\end{matrix}
} =
\&\left(
\boxed
{
\begin{matrix}
0\\
1\\
0
\end{matrix}
}(x),
\boxed
{
\begin{matrix}
0\\
1\\
0
\end{matrix}
}(y)
\right), \quad
\boxed
{
\begin{matrix}
0 & 1 & 1\\
1 & 1 & 1\\
1 & 1 & 1
\end{matrix}
}=
\vee\left(
\boxed
{
\begin{matrix}
0\\
1\\
1
\end{matrix}
}(x),
\boxed
{
\begin{matrix}
0\\
1\\
1
\end{matrix}
}(y)
\right).
$$
\end{proof}

В силу ряда причин доказательство для классов 
$\Sigma^{\{0,1\}\{2\}}_{T_0},\Sigma^{\{0,1\}\{2\}}_{T_1}$
в некоторых моментах отличается от доказательства теоремы \ref{TKDL}

\begin{theorem}\label{TT0}
Класс функций $\Sigma^{\{0,1\}\{2\}}_{T_0}$, 
а также двойственные ему классы неявно предполны в $P_3$.
\end{theorem}
\begin{proof}
Выпишем все одноместные функции класса
$\Sigma^{\{0,1\}\{2\}}_{T_0}$.
$$
\boxed
{
\begin{matrix}
0\\
0\\
0
\end{matrix}
}
\boxed
{
\begin{matrix}
0\\
0\\
2
\end{matrix}
}
\boxed
{
\begin{matrix}
0\\
1\\
0
\end{matrix}
}
\boxed
{
\begin{matrix}
0\\
1\\
2
\end{matrix}
}
\boxed
{
\begin{matrix}
2\\
2\\
0
\end{matrix}
}
\boxed
{
\begin{matrix}
2\\
2\\
2
\end{matrix}
}
$$

Неявная неполнота класса $\Sigma^{\{0,1\}\{2\}}_{T_0}$ следует из 
неявной неполноты класса 
$T_0\subset P_2$ и леммы \ref{Unotfull}.

Докажем, что при добавлении к $\Sigma^{\{0,1\}\{2\}}_{T_0}$ 
любой функции $f$ не из этого класса система становится неявно полной.


Сначала рассмотрим случай, когда $f\in U_{\{0,1\}\{2\}}$.

Если функция $f$ сохраняет разбиение, 
но не принадлежит $\Sigma^{\{0,1\}\{2\}}_{T_0}$, 
то ее булево ограничение на некотором 
блоке $B$ не лежит в классе $T_0\subset P_2$. Подставив константу $2$
в функцию $f$ на те места, 
где у наборов из $B$ стоит двойка, получим функцию $f'$, у которой 
булево ограничение на главном блоке не лежит в классе $T_0$.

Рассмотрим булево ограничение $\hat{f'}$ функции $f'$ на главный блок.
По построению $\hat{f'}\notin T_0$. Из того, что 
$[T_0\cup \{\hat{f'}\}] = P_2$, и 
следствия из леммы \ref{Ubool} следует, что в 
$\left[\Sigma^{\{0,1\}\{2\}}_{T_0}\cup \{f'\}\right]$ найдется функция 
с любым булевым ограничением. 
В частности, там найдется пара функций вида
$$
\&(x,y)=
\boxed
{
\begin{matrix}
0 & 0 & *\\
0 & 1 & *\\
* & * & *
\end{matrix}
}\quad
\vee(x,y)=
\boxed
{
\begin{matrix}
0 & 1 & *\\
1 & 1 & *\\
* & * & *
\end{matrix}
},
$$
из которых с помощью функций 
$
\boxed
{
\begin{matrix}
0\\
1\\
0
\end{matrix}
}
$ и 
$
\boxed
{
\begin{matrix}
0\\
1\\
1
\end{matrix}
}
$, лежащих в $\Sigma^{\{0,1\}\{2\}}_{T_0}$, 
так же, как и в доказательстве предыдущей теоремы, 
нетрудно построить неявно полную систему $S_1$.

Для доказательства теоремы для случая, когда 
$f\notin U_{\{0,1\}\{2\}}$, нам понадобится следующая 
\begin{lemma}\label{U1by2const}
Из функции $f(x_1, x_2, \dots, x_n)\notin U_{\{0,1\}\{2\}}$ 
и констант $0$ и $2$ можно получить некоторую одноместную функцию 
$g(x)\notin U_{\{0,1\}\{2\}}$.
\end{lemma}
\begin{proof}
Так как $f\notin U_{\{0,1\}\{2\}}$, то найдутся два набора 
$\tilde{\alpha}$ и $\tilde{\beta}$ из одного блока такие, 
что $f(\tilde{\alpha})=2, f(\tilde{\beta})\ne 2$. 
После выделения этого блока константами 2 получим функцию 
$f'(x_1,\dots,x_m), m\le n$. 
Разобьем булев куб $B_2^m$ в прямую сумму двух множеств. 
К первому множеству отнесем те наборы, 
на которых функция $f'$ принимает значение 2, 
а ко второму остальные. Тогда найдутся два соседних набора, пусть 
$\tilde{\alpha}'=(0,\dots,0,0,1,\dots,1)$ и 
$\tilde{\beta}'=(0,\dots,0,1,1,\dots,1)$, лежащие в разных множествах. 
Пусть $f'(\tilde{\alpha}')=2, f'(\tilde{\beta}')=c$, где $c\in \{0,1\}$. 
Возьмем $f''(x,y)=f'(0,\dots,0,x,y,\dots,y)$. 
$f''(x,y)$ имеет вид 
$
\boxed
{
\begin{matrix}
* & 2 & *\\
* & c & *\\
* & * & *
\end{matrix}
}
$.
Если $f''(0,0)=2$, то $g(x)=f''(x,x)=
\boxed
{
\begin{matrix}
2\\
c\\
*
\end{matrix}
}
$. Иначе $g(x)=f''(0,x)=
\boxed
{
\begin{matrix}
d\\
2 \\
*
\end{matrix}
}
$, где $d=f'(0,0)\in \{0,1\}$. 
В любом случае получаем одноместную функцию не из 
$U_{\{0,1\}\{2\}}$.
\end{proof}

Таким образом, из функции $f$ можно получить одноместную функцию 
$g(x)\notin U_{\{0,1\}\{2\}}$. 
Она имеет вид 
$
\boxed
{
\begin{matrix}
a\\
2\\
*
\end{matrix}
}
$ или 
$
\boxed
{
\begin{matrix}
2\\
a\\
*
\end{matrix}
}
$, где $a\in \{0,1\}$.

Покажем, что в любом случае мы сможем по суперпозиции 
выразить неявно полную систему 
$$
\boxed
{
\begin{matrix}
0 & 0 & 2\\
0 & 0 & 2\\
2 & 2 & 2
\end{matrix}
}
\boxed
{
\begin{matrix}
0 & 0 & 0\\
0 & 0 & 0\\
0 & 0 & 2
\end{matrix}
}
\boxed
{
\begin{matrix}
0\\
0\\
0
\end{matrix}
}
\boxed
{
\begin{matrix}
2\\
0\\
2
\end{matrix}
}
,
$$
двойственную системе $S_2$ относительно подстановки $(021)$.
Так как первые три функции системы принадлежат классу
$\Sigma^{\{0,1\}\{2\}}_{T_0}$, 
для доказательства теоремы достаточно получить функцию
$
\boxed
{
\begin{matrix}
2\\
0\\
2
\end{matrix}
}
$.

Рассмотрим функцию 
$
g'(x)=(
\boxed
{
\begin{matrix}
0\\
0\\
2
\end{matrix}
}(g(x))
$. Она совпадает с одной из четырех функций
$
\boxed
{
\begin{matrix}
0\\
2\\
0
\end{matrix}
}
\boxed
{
\begin{matrix}
0\\
2\\
2
\end{matrix}
}
\boxed
{
\begin{matrix}
2\\
0\\
0
\end{matrix}
}
\boxed
{
\begin{matrix}
2\\
0\\
2
\end{matrix}
}
$. 
Каждая из первых трех функций вместе с функциями из 
$\Sigma^{\{0,1\}\{2\}}_{T_0}$ порождает по суперпозиции функцию 
$
\boxed
{
\begin{matrix}
2\\
0\\
2
\end{matrix}
}
$. Это следует из соотношений

$$
\boxed
{
\begin{matrix}
2\\
2\\
0
\end{matrix}
}
\left(
\boxed
{
\begin{matrix}
0\\
2\\
0
\end{matrix}
}
\right)=
\boxed
{
\begin{matrix}
2\\
0\\
2
\end{matrix}
};\quad
\boxed
{
\begin{matrix}
0\\
2\\
2
\end{matrix}
}
\left(
\boxed
{
\begin{matrix}
0\\
1\\
0
\end{matrix}
}
\right)=
\boxed
{
\begin{matrix}
0\\
2\\
0
\end{matrix}
};\quad
\boxed
{
\begin{matrix}
2\\
0\\
0
\end{matrix}
}
\left(
\boxed
{
\begin{matrix}
2\\
0\\
0
\end{matrix}
}
\right)=
\boxed
{
\begin{matrix}
0\\
2\\
2
\end{matrix}
}.
$$
Здесь мы или явно показываем, как получить функцию
$
\boxed
{
\begin{matrix}
2\\
0\\
2
\end{matrix}
}
$, или сводим к уже разобранным случаям.

Таким образом, какой бы ни была функция $f$, система 
$\Sigma^{\{0,1\}\{2\}}_{T_0}\cup \{f\}$ неявно полна в $P_3$, и, 
следовательно, 
класс $\Sigma^{\{0,1\}\{2\}}_{T_0}$ 
является неявно предполным.
 
\end{proof}

\section{Полнота описания}
Докажем, 
что список неявно предполных классов в $P_3$ исчерпывается
классами, описанными выше. 
Для этого докажем, 
что для каждого $m=0,1,2,3$ не существует неявно предполных классов
с $m$ константами отличных от описанных классов.

А именно, 
рассмотрим все возможные 1-основания, 
содержащие тождественную функцию,
с заданным числом констант.
Для некоторых из них покажем, 
что любой класс с таким основанием содержится в некотором 
описанном неявно предполном классе.
Для других покажем, 
что классы с такими основаниями неявно предполные 
тогда и только тогда, 
когда совпадают с одним из классов семейства $\mathfrak{A}$.
Наконец, для оставшихся оснований покажем, 
что если класс с таким основанием предполон, 
то он должен содержать несколько функций,
которые вместе с рассматриваемыми основаниями порождают 
неявно полный класс.
\subsection{Классы без констант}
\begin{theorem}\label{all0const}
Существует ровно 4 неявно предполных класса в $P_3$,
не содержащих констант:
$S, \Sigma_S^{\{0,1\}\{2\}},\Sigma_S^{\{0,2\}\{1\}},
\Sigma_S^{\{1,2\}\{0\}}$
\end{theorem}

Можно показать, 
что любой замкнутый по суперпозиции класс одноместных функций,
содержащий селектор и не содержащий констант, 
совпадает с одним из следующих классов или двойственен ему:
\begin{gather*}
1.1-1.5)\quad
\boxed
{\begin{matrix}0\\1\\2\end{matrix}};\quad
\boxed{\begin{matrix}0\\1\\2\end{matrix}}
\boxed{\begin{matrix}0\\2\\1\end{matrix}};\quad
\boxed{\begin{matrix}0\\1\\0\end{matrix}}
\boxed{\begin{matrix}0\\1\\2\end{matrix}};\quad
\boxed{\begin{matrix}0\\1\\0\end{matrix}}
\boxed{\begin{matrix}0\\1\\1\end{matrix}}
\boxed{\begin{matrix}0\\1\\2\end{matrix}};\quad
\boxed{\begin{matrix}0\\1\\0\end{matrix}}
\boxed{\begin{matrix}0\\1\\2\end{matrix}}
\boxed{\begin{matrix}2\\1\\2\end{matrix}};\quad
\boxed{\begin{matrix}0\\0\\2\end{matrix}}
\boxed{\begin{matrix}0\\1\\2\end{matrix}}
\boxed{\begin{matrix}1\\0\\2\end{matrix}}
\boxed{\begin{matrix}1\\1\\2\end{matrix}};\quad
\\
2.1)\quad
\boxed{\begin{matrix}0\\1\\2\end{matrix}}
\boxed{\begin{matrix}1\\2\\0\end{matrix}}
\boxed{\begin{matrix}2\\0\\1\end{matrix}};\quad
\\
3.1-3.3)\quad
\boxed{\begin{matrix}0\\1\\0\end{matrix}}
\boxed{\begin{matrix}0\\1\\1\end{matrix}}
\boxed{\begin{matrix}0\\1\\2\end{matrix}}
\boxed{\begin{matrix}1\\0\\0\end{matrix}}
\boxed{\begin{matrix}1\\0\\1\end{matrix}}
\boxed{\begin{matrix}1\\0\\2\end{matrix}};\quad
\boxed{\begin{matrix}0\\1\\0\end{matrix}}
\boxed{\begin{matrix}0\\1\\1\end{matrix}}
\boxed{\begin{matrix}0\\1\\2\end{matrix}}
\boxed{\begin{matrix}1\\0\\0\end{matrix}}
\boxed{\begin{matrix}1\\0\\1\end{matrix}};\quad
\boxed{\begin{matrix}0\\1\\0\end{matrix}}
\boxed{\begin{matrix}0\\1\\2\end{matrix}}
\boxed{\begin{matrix}1\\0\\1\end{matrix}};\quad
\\
4.1)\quad
\boxed{\begin{matrix}0\\1\\2\end{matrix}}
\boxed{\begin{matrix}0\\2\\1\end{matrix}}
\boxed{\begin{matrix}1\\0\\2\end{matrix}}
\boxed{\begin{matrix}1\\2\\0\end{matrix}}
\boxed{\begin{matrix}2\\0\\1\end{matrix}}
\boxed{\begin{matrix}2\\1\\0\end{matrix}};
\\
5.1-5.2)\quad
\boxed{\begin{matrix}0\\0\\2\end{matrix}}
\boxed{\begin{matrix}0\\1\\2\end{matrix}}
\boxed{\begin{matrix}1\\1\\2\end{matrix}}
\boxed{\begin{matrix}2\\2\\0\end{matrix}}
\boxed{\begin{matrix}2\\2\\1\end{matrix}};\quad
\boxed{\begin{matrix}0\\0\\2\end{matrix}}
\boxed{\begin{matrix}0\\1\\2\end{matrix}}
\boxed{\begin{matrix}1\\0\\2\end{matrix}}
\boxed{\begin{matrix}1\\1\\2\end{matrix}}
\boxed{\begin{matrix}2\\2\\0\end{matrix}}
\boxed{\begin{matrix}2\\2\\1\end{matrix}}.
\end{gather*}

Каждое из оснований 1.1 -- 1.5 состоит из функций, 
сохраняющих некоторую константу.
Следовательно, классы сохранения этих оснований 
содержатся в классе сохранения соответствующей константы
и не являются неявно предполными.

Вычеркивая из матрицы
$
\begin{pmatrix}
0 & 1 & 2\\
1 & 2 & 0\\
2 & 0 & 1
\end{pmatrix}
$, соответствующей основанию 2.1,
последнюю строку, получим матрицу 
$
\begin{pmatrix}
0 & 1 & 2\\
1 & 2 & 0\\
\end{pmatrix}
$, которая определяет класс $S$.
Таким образом, по утверждению~\ref{matrix}
любой класс с основанием 2.1 содержится в 
неявно предполном классе $S$.

Рассмотрим матрицы, соответствующие основаниям 3.1 -- 3.3.
Вычеркивая из них
последнюю строку и убирая повторяющиеся столбцы, получим матрицу 
$
\begin{pmatrix}
0 & 1\\
1 & 0\\
\end{pmatrix}
$, которая определяет класс $\Sigma_S^{\{0,1\}}$.
Таким образом, по утверждению~\ref{matrix}
любой класс с основаниями 3.1 -- 3.3 содержится 
в неявно предполном классе $\Sigma_S^{\{0,1\}}$.

\begin{statement}
Максимальный надкласс основания 1.4 содержится в классе
$\mathfrak{N}$.
\end{statement}
\begin{proof}
Пусть существует функция $f(\tilde{x})$, 
сохраняющая основание 1.4 и 
выделяющая вершину некоторого квадрата. Тогда 
\begin{gather*}
\tilde{\delta_1}=(\alpha_1,\beta_1,\gamma_1,\dots,\gamma_n),\; 
f(\tilde{\delta_1})=d_1\\
\tilde{\delta_2}=(\alpha_2,\beta_2,\gamma_1,\dots,\gamma_n), \;
f(\tilde{\delta_2})=d_2\\
\tilde{\delta_3}=(\alpha_2,\beta_1,\gamma_1,\dots,\gamma_n), \;
f(\tilde{\delta_3})=d_3\\
\tilde{\delta_4}=(\alpha_1,\beta_2,\gamma_1,\dots,\gamma_n), \;
f(\tilde{\delta_4})=d_4,
\end{gather*} 
где $\alpha_1\neq \alpha_2, \beta_1\neq \beta_2$.
Кроме того, как минимум одно из значений $d_i$ отлично от всех остальных.

Без ограничения общности 
$\gamma_1,\dots,\gamma_n=(0,\dots,0,1,\dots,1,2,\dots,2)$.
Рассмотрим функцию
$$
g(x,y,z)=f(x,y,z,\dots,z,
\boxed
{
\begin{matrix}
1\\
0\\
2
\end{matrix}
}(z),\dots,
\boxed
{
\begin{matrix}
1\\
0\\
2
\end{matrix}
}(z),
\boxed
{
\begin{matrix}
2\\
0\\
1
\end{matrix}
}(z),\dots,
\boxed
{
\begin{matrix}
2\\
0\\
1
\end{matrix}
}(z)).
$$
Она выделяет квадрат, задаваемый вершинами $(\alpha_1,\beta_1,0)$ и 
$(\alpha_2,\beta_2,0)$.

Если взять два набора, различающихся в каждой компоненте, 
то и значения функции $g$ на этих наборах должны быть различными. 
Иначе подстановкой одноместных функций из основания, 
а также отождествлением переменных можно будет получить 
одноместную функцию, которая принимает некоторое значение дважды. 
А таких функций в основании нет. По тем же причинам, 
если взять фиксированный набор $\tilde{\omega}$, 
то значение функции на нем отлично от значений 
на всех наборах, которые отличаются от $\tilde{\omega}$ во всех компонентах. 

В частности, на наборе $\tilde{\delta_0}=(\alpha_3,\beta_3,2)$
значение $g(\tilde{\delta_0})=d_0$
должно быть отличным от всех $d_i$ при $i=1,\dots,4$.
Отсюда следует, что среди $d_i$ при $i=1,\dots,4$ есть всего два значения, 
одно из которых принимается лишь однажды. Будем считать, 
что $d_1=d_3=d_4\neq d_2$. Определим значения функции на некоторых 
других наборах.
\begin{gather*}
g(\alpha_2,\beta_2,0)=d_2,\;
g(\alpha_1,\beta_2,0)=d_1 \Rightarrow
g(\alpha_3,\beta_1,2)=d_0\\
g(\alpha_2,\beta_2,0)=d_2,\;
g(\alpha_2,\beta_1,0)=d_1 \Rightarrow
g(\alpha_1,\beta_3,1)=d_1\\
g(\alpha_1,\beta_2,0)=d_1,\;
g(\alpha_3,\beta_1,2)=d_0 \Rightarrow
g(\alpha_2,\beta_3,1)=d_2\\
g(\alpha_2,\beta_1,0)=d_1,\;
g(\alpha_1,\beta_3,1)=d_0 \Rightarrow
g(\alpha_3,\beta_2,2)=d_2
\end{gather*}
Мы получили одинаковое значение на наборах $(\alpha_2,\beta_3,1)$ и 
$(\alpha_3,\beta_2,2)$, которые различны во всех компонентах. 
А такого быть не может.

Следовательно, наше предположение неверно, и максимальный надкласс 
основания 1.4 действительно содержится в классе квазилинейных функций.
\end{proof}

Так как все классы с основанием 1.4 содержатся в $\mathfrak{N}$ 
и не совпадают с ним, не существует неявно предполных классов 
с таким основанием.

\begin{lemma}\label{Umax}
Если в основании явно замкнутого класса $\mathfrak{A}$ 
найдется тройка функций вида
$
\boxed
{
\begin{matrix}
0\\
0\\
*
\end{matrix}
}
\boxed
{
\begin{matrix}
1\\
1\\
*
\end{matrix}
}
\boxed
{
\begin{matrix}
2\\
2\\
*
\end{matrix}
}
$, 
а само основание целиком содержится в классе $U_{\{0,1\}\{2\}}$, 
то максимальный надкласс этого основания 
также содержится в $U_{\{0,1\}\{2\}}$.
\end{lemma}
\begin{proof}
Пусть в $\mathfrak{A}$ существует функция 
$f(\tilde{x})\notin U_{\{0,1\}\{2\}}$. Тогда найдутся два набора
$\tilde{\alpha_1}, \tilde{\alpha_2}$, без ограничения общности имеющие вид 
$\tilde{\alpha_i}=(\tilde{\beta_i},\tilde{2})$, где $\tilde{\beta_i}\in E_2^m$ 
такие, что
$$
\left[
\begin{aligned}
\left\lbrace
\begin{aligned}
f(\tilde{\alpha_1})=2\\
f(\tilde{\alpha_2})\neq 2
\end{aligned}
\right.\\
\left\lbrace
\begin{aligned}
f(\tilde{\alpha_1})\neq 2\\
f(\tilde{\alpha_2})=2
\end{aligned}
\right.
\end{aligned}
\right.
.
$$
Причем эти наборы можно подобрать таким образом, чтобы наборы 
$\tilde{\beta_1}$ и $\tilde{\beta_2}$ различались только в одной компоненте.

Подставим в функцию $f$ столбцы
$
\boxed
{
\begin{matrix}
0\\
0\\
*
\end{matrix}
}
\boxed
{
\begin{matrix}
0\\
1\\
2
\end{matrix}
}
\boxed
{
\begin{matrix}
1\\
1\\
*
\end{matrix}
}
\boxed
{
\begin{matrix}
2\\
2\\
*
\end{matrix}
}
$
таким образом, чтобы в верхней строке 
реализовывался, меньший из наборов $\tilde{\alpha_i}$, а во второй --- больший. 
Тогда в результирующем столбце одно из значений в верхних двух строчках будет 
равно $2$, а второе принадлежать $\{0,1\}$. Но функций, 
соответствующих таким столбцам, не может быть в основаниях, удовлетворяющих 
условию леммы.

Противоречие.
\end{proof}

\begin{note}
Очевидно, что в $U_{\{0,1\}\{2\}}$ будет содержаться не только 
максимальный надкласс, но и всякий класс с таким основанием.
\end{note}

\begin{lemma}\label{Udizkon}
Пусть $\mathfrak{A}$ --- неявно предполный класс, 
содержащий тройку функций вида 
$
\boxed
{
\begin{matrix}
0\\
0\\
*
\end{matrix}
}
\boxed
{
\begin{matrix}
1\\
1\\
*
\end{matrix}
}
\boxed
{
\begin{matrix}
2\\
2\\
*
\end{matrix}
}
$, $\mathfrak{A}\subseteq U_{\{0,1\}\{2\}}$, 
причем 
$\mathfrak{A}\neq \Sigma_{\mathfrak{B}}^{\{a,b\}\{c\}}$,
где $\mathfrak{B}\in \{T_0, T_1, L, K, D\}$.
Тогда в $\mathfrak{A}$ есть функции вида
$
\vee(x,y)=
\boxed
{
\begin{matrix}
0 & 1 & *\\
1 & 1 & *\\
* & * & *
\end{matrix}
}
$ и
$
\&(x,y)=
\boxed
{
\begin{matrix}
0 & 0 & *\\
0 & 1 & *\\
* & * & *
\end{matrix}
}
$.
\end{lemma}

\begin{proof}
В силу следствия из леммы \ref{Ubool} 
лемму можно переформулировать в терминах классов в $P_2$:

Пусть $\mathfrak{B}\subset P_2$, $0,1\in \mathfrak{B}$, причем
$\mathfrak{B}$ не содержится ни в одном из классов $T_0,T_1,L,K,D$. 
Тогда функции $\vee, \&$ принадлежат $\mathfrak{B}$.

Такая система должна содержать в себе класс $M$ и, следовательно, 
функции $\vee$ и $\&$.	
\end{proof}

\begin{statement}
Не существует неявно предполных классов в $P_3$ с основаниями 
5.1 и 5.2.
\end{statement}
\begin{proof}

Предположим, что существует неявно предполный класс 
с основанием 5.1 или 5.2. 
Тогда он должен содержать функции $\vee$ и $\&$.
Покажем, что в явном замыкании функций типа $\vee$ и $\&$, 
а также функций из основания 5.1, 
лежит неявно полная система 
$
S_6=
\boxed
{
\begin{matrix}
0 & 0 & 2\\
0 & 1 & 2\\
0 & 0 & 2
\end{matrix}
}
\boxed
{
\begin{matrix}
0 & 1 & 2\\
1 & 1 & 2\\
0 & 0 & 2
\end{matrix}
}
\boxed
{
\begin{matrix}
2\\
2\\
0
\end{matrix}
}
$.

Так как все функции должны принадлежать классу $U_{\{0,1\}\{2\}}$,
то значения функции $\vee$ на блоках $(*,2)$ и $(2,*)$ 
эквивалентны относительно разбиения $\{0,1\}\{2\}$. При этом значения на одном
блоке не эквивалентны значениям на другом, так как в таком случае можно было бы
выразить одноместную функцию 
$
\vee
\left(
x,
\boxed
{
\begin{matrix}
2\\
2\\
0
\end{matrix}
}(x)
\right)
$, которая не принадлежит ни одному из оснований 5.1, 5.2.

Так как $\vee(x,x)$ и $\&(x,x)$ 
должны принадлежать нашему основанию, то
они должны совпадать с 
$
\boxed
{
\begin{matrix}
0\\
1\\
2
\end{matrix}
}
$. Поэтому
без ограничения общности можно считать, что функция $\vee$ имеет вид
$
\boxed
{
\begin{matrix}
0   & 1   & 2\\
1   & 1   & 2\\
a_1 & b_1 & 2
\end{matrix}
}
$, а функция $\&$ --- 
$
\boxed
{
\begin{matrix}
0   & 0   & 2\\
0   & 1   & 2\\
a_2 & b_2 & 2
\end{matrix}
}
$
где $a_i,b_i\in \{0,1\}$.
Можно показать, что вне зависимости от значений $a_1,a_2,b_1,b_2$
из функций $\vee, \&$, а также функций
$
\boxed{
\begin{matrix}
0\\0\\2
\end{matrix}
}
\boxed{
\begin{matrix}
1\\1\\2
\end{matrix}
}
$ суперпозицией можно получить функции
$
\boxed
{
\begin{matrix}
0 & 0 & 2\\
0 & 1 & 2\\
0 & 0 & 2
\end{matrix}
}
\boxed
{
\begin{matrix}
0 & 1 & 2\\
1 & 1 & 2\\
0 & 0 & 2
\end{matrix}
}
$

Таким образом, мы явно выразили неявно полный класс, 
а значит $\mathfrak{A}$ 
не был неявно предполным. 
Противоречие.
\end{proof}

Все возможные основания без констант рассмотрены. 
Теорема~\ref{all0const} доказана.

\subsection{Классы с одной константой}
\begin{theorem}\label{all1const}
Существует ровно 3 неявно предполных класса в $P_3$, содержащих в точности 
одну константу: $T_{\{0\},0},T_{\{1\},0},T_{\{2\},0}$.
\end{theorem}

\begin{proof}
Предположим, что явно замкнутый класс $F$ содержит ровно одну 
константу $a$.
Тогда все функции $F$ принадлежат классу $T_{\{a\},0}$.

Действительно, пусть некоторая функция $f(\tilde{x})$ нашего класса 
не сохраняет значение $a$. 
Тогда $f(a,a,\dots,a)\neq a$, и мы выразили некоторую константу отличную от $a$. 
Противоречие.

Таким образом, каждый явно замкнутый класс с одной константой содержится в одном из неявно 
предполных классов $T_{\{0\},0},T_{\{1\},0},T_{\{2\},0}$.
\end{proof}

\subsection{Классы с двумя константами}
\begin{theorem}\label{all2const}
Существует ровно 18 неявно предполных классов в $P_3$, содержащих 
в точности две константы: 
$\Sigma_K^{\{0,1\}}$, $\Sigma_K^{\{0,2\}}$, $\Sigma_K^{\{1,2\}}$,
$\Sigma_D^{\{0,1\}}$, $\Sigma_D^{\{0,2\}}$, $\Sigma_D^{\{1,2\}}$,
$\Sigma_L^{\{0,1\}}$, $\Sigma_L^{\{0,2\}}$, $\Sigma_L^{\{1,2\}}$, 
$\Sigma_{T_0}^{\{0,1\}\{2\}}$, $\Sigma_{T_0}^{\{0,2\}\{1\}}$,
$\Sigma_{T_0}^{\{1,2\}\{0\}}$, $\Sigma_{T_1}^{\{0,1\}\{2\}}$,
$\Sigma_{T_1}^{\{0,2\}\{1\}}$, $\Sigma_{T_1}^{\{1,2\}\{0\}}$,
$Y_0$, $Y_1$, $Y_2$.
\end{theorem}

Можно показать, 
что любой замкнутый по суперпозиции класс одноместных функций,
содержащий селектор, константы $0$, $1$ 
и не содержащий константу $2$, 
получается из одной из следующих систем (или двойственной к ней)
добавлением функций $0,1,x$:

\begin{gather*}
1.1-1.13)\\
\soio\soii;\quad 
\sooi\soio\soii;\quad 
\sooi\sooz\soio\soii;\quad 
\sooi\soio\soii\siio;\quad 
\sooi\sooz\soio\soii\siio;\quad 
\sooi\sooz\soio\soii\siio\siiz;\quad 
\\
\soio\soii\sioo\sioi;\quad
\sooi\soio\soii\sioo\sioi\siio;\quad
\sooi\sooz\soio\soii\sioo\sioi\siio;\quad
\sooi\sooz\soio\soii\sioo\sioi\siio\siiz;\quad
\\
\soio\soii\sioo\sioi\sioz;\quad
\sooi\soio\soii\sioo\sioi\sioz\siio;\quad
\sooi\sooz\soio\soii\sioo\sioi\sioz\siio\siiz;\quad
\\
2.1-2.6)\\
\sooi\soio;\quad
\sooi\sooz\soio;\quad
\sooi\soio\sioi\siio;\quad
\sooi\sooz\soio\sioi\siio;\quad
\sooi\soio\sioi\siio\siiz;\quad
\sooi\sooz\soio\sioi\siio\siiz;\quad
\\
3.1-3.8)\\
\soio\siio;\quad
\soio\siio\siiz;\quad
\sooi\soio\siio;\quad
\sooz\soio\siio;\quad
\sooi\sooz\soio\siio;\quad
\sooi\soio\siio\siiz;\quad
\sooz\soio\siio\siiz;\quad
\sooi\sooz\soio\siio\siiz;\quad
\end{gather*}
\begin{gather*}
4.1-4.11)\\
\sooi;\quad
\sooi\sooz;\quad
\sooi\siio;\quad
\sooi\siiz;\quad
\sooi\siio;\quad
\sooi\sooz\siio;\quad
\sooi\sooz\siio;\quad
\sooi\sooz\siiz;\quad
\\
\sooi\sooz\siio\siiz;\quad
\sooi\sioz\siio;\quad
\sooi\sooz\sioz\siio\siiz;\quad
\\
5.1-5.4)\\
\varnothing;\quad
\sooz;\quad
\soio;\quad
\soio\sooz;\quad
\\
6.1-6.3)\\
\sioz;\quad
\sooz\siiz;\quad
\sooz\sioz\siiz;\quad
\end{gather*}

Докажем, 
что среди классов с такими основаниями неявно предполными могут 
быть только указанные в теореме.

\begin{lemma}\label{Tdizkon}
Пусть $\mathfrak{A}$ --- 
неявно предполный класс, содержащий пару функций вида 
$
\boxed
{
\begin{matrix}
0\\
0\\
*
\end{matrix}
}
\boxed
{
\begin{matrix}
1\\
1\\
*
\end{matrix}
}
$, $\mathfrak{A}\subseteq T_{\{0,1\},0}$, причем 
$\mathfrak{A}\notin
\{\Sigma_{S}^{\{0,1\}},\Sigma_L^{\{0,1\}},\Sigma_{K}^{\{0,1\}},
\Sigma_{D}^{\{0,1\}}\}$. 
Тогда в $\mathfrak{A}$ есть функции вида
$
\vee(x,y)=
\boxed
{
\begin{matrix}
0 & 1 & *\\
1 & 1 & *\\
* & * & *
\end{matrix}
}
$ и
$
\&(x,y)=
\boxed
{
\begin{matrix}
0 & 0 & *\\
0 & 1 & *\\
* & * & *
\end{matrix}
}
$.
\end{lemma}
\begin{proof}
Рассмотрим булево ограничение $\mathfrak{A}$.
Оно не может содержаться в классах $S,L,K,D$, 
так как в таком случае $\mathfrak{A}$ будет содержаться 
в соответствующих неявно предполных классах.
Кроме того, оно не содержится в классах $T_1$ и $T_0$, 
так как содержит функции
$
\boxed
{
\begin{matrix}
0\\
0\\
*
\end{matrix}
}
\boxed
{
\begin{matrix}
1\\
1\\
*
\end{matrix}
}$.
Из рассмотрения решетки Поста получается, 
что булево ограничение $\mathfrak{A}$ 
содержит класс монотонных функций и, в частности, 
функции $\lor(x,y)$ и $\&(x,y)$,
из чего следует утверждение леммы.
\end{proof}

\begin{lemma}\label{2const1}
Пусть явно замкнутый класс $\mathfrak{A}\subseteq P_3$ 
содержит константы 
$0$ и $1$, а также функции вида $\vee(x,y)$ и $\&(x,y)$. Тогда:

а) 
$
\sooi,
\soio
\in \mathfrak{A}
\Rightarrow
\soii
\in \mathfrak{A}.
$

б)
$
\soio,
\soii
\in \mathfrak{A}
\Rightarrow
\mathfrak{A}
$ неявно полон.

в) $
\soio,
\siio
\in \mathfrak{A}
\Rightarrow
\mathfrak{A}
$ неявно полон.
\end{lemma}

\begin{proof}
а) 
$
\vee
\left(
\sooi(x),
\soio(x)
\right)
=
\soii
$.

В пунктах б) и в) из имеющихся функций 
несложно суперпозицией получить системы
$$
S_1=
\sooo
\siii
\boxed
{
\begin{matrix}
0 & 0 & 0\\
0 & 1 & 0\\
0 & 0 & 0
\end{matrix}
}
\boxed
{
\begin{matrix}
0 & 1 & 1\\
1 & 1 & 1\\
1 & 1 & 1
\end{matrix}
};\quad
\sooo
\siio
\boxed
{
\begin{matrix}
0 & 0 & 0\\
0 & 1 & 0\\
0 & 0 & 0
\end{matrix}
}
\boxed
{
\begin{matrix}
0 & 1 & 0\\
1 & 1 & 1\\
0 & 1 & 0
\end{matrix}
},
$$
порождающие неявно полные классы Ореховой.
\end{proof}

\begin{lemma}\label{2const2}
Пусть явно замкнутый класс $\mathfrak{A}\subseteq P_3$ 
содержит константы 
$0$ и $1$, функции вида $\vee(x,y)$ и $\&(x,y)$,а также функцию
$\sioz$. 
Тогда в основании должна лежать хотя бы одна из функций 
$\soio\soii\siiz$.
\end{lemma}
\begin{proof}
Пусть в основании нет ни одной из указанных функций. 
Тогда $\vee(x,x)=\&(x,x)=x$,
и $\vee\left(x, \sioz(x) \right)=\siiz$, 
которая не принадлежит нашему основанию. 
Противоречие.
\end{proof}

\begin{lemma}\label{2const3}
Пусть явно замкнутый класс $\mathfrak{A}\subseteq P_3$ 
содержит константы 
$0$ и $1$, функции 
$\sooz$ и $\siiz$, а также 
функции вида $\vee(x,y)$ и $\&(x,y)$,  
но не содержит функции
$\soio$ и $\soii$.
Тогда класс $\mathfrak{A}$ содержит пару функций 
$$
max(x,y)=
\boxed
{
\begin{matrix}
0 & 1 & 2\\
1 & 1 & 2\\
2 & 2 & 2
\end{matrix}
},\;
min_{01}(x,y)=
\boxed
{
\begin{matrix}
0 & 0 & 2\\
0 & 1 & 2\\
2 & 2 & 2
\end{matrix}
}.
$$
\end{lemma}

\begin{proof}
Действительно, так как функции
$\soio$ и $\soii$ не принадлежат классу, то
$\vee(x,x)=\vee(0,x)=\vee(x,0)=\&(x,x)=\&(1,x)=\&(x,1)=x$, 
а значит
$$
\vee(x,y)=
\boxed
{
\begin{matrix}
0 & 1 & 2\\
1 & 1 & a_1\\
2 & a_2 & 2
\end{matrix}
};\,
\&(x,y)=
\boxed
{
\begin{matrix}
0 & 0 & b_1\\
0 & 1 & 2\\
b_2 & 2 & 2
\end{matrix}
}.
$$

С помощью суперпозиции несложно из полученных функций,
а также функций $\sooz,\siiz$ получить 
требуемые $max(x,y)$ и $min_{01}(x,y)$.
\end{proof}

\begin{lemma}\label{2const32}
Пусть явно замкнутый класс $\mathfrak{A}\subseteq P_3$ 
содержит константы $0$ и $1$, функцию $\sooi$, а также 
функции вида $\vee(x,y)$ и $\&(x,y)$,  
но не содержит функции $\soio$ и $\soii$.
Тогда класс $\mathfrak{A}$ содержит пару функций $\sooz\siiz$.
\end{lemma}
\begin{proof}
Так же, как и в доказательстве предыдущей леммы, 
можно утверждать, что
$$
\vee(x,y)=
\boxed
{
\begin{matrix}
0 & 1 & 2\\
1 & 1 & a_1\\
2 & a_2 & 2
\end{matrix}
};\,
\&(x,y)=
\boxed
{
\begin{matrix}
0 & 0 & b_1\\
0 & 1 & 2\\
b_2 & 2 & 2
\end{matrix}
}.
$$
Легко видеть, 
что 
$\&\left(\sooi(x),x \right) = \sooz$. 
Кроме того, так как
$$
\vee\left(
\sooi(x),x \right) = 
\vee\left(x,\sooi(x) \right) = x,
$$ то функция $\vee(x,y)$ равна 
$
\boxed
{
\begin{matrix}
0 & 1 & 2\\
1 & 1 & 2\\
2 & 2 & 2
\end{matrix}
}
$,
а значит, что $\vee(1,y) = \siiz$.
\end{proof}

\begin{statement}\label{2const33}
Пусть система функций $\mathfrak{A}\subseteq P_3$
содержит функции $0,1,max(x,y), min_{01}(x,y), f(x)$, 
где $f(x)$~--- одна из функций 
$\sooi\siio\szzo$.
Тогда $\mathfrak{A}$ неявно полна в $P_3$.
\end{statement}
\begin{proof}
Для каждой из указанных в условии функций $f(x)$ 
система $\{0,1,max(x,y), min_{01}(x,y), f(x)\}$
порождает один из классов Ореховой.
\end{proof}

\begin{lemma}\label{2const4}
Если в основании явно замкнутого класса лежат константы $0,1$, 
но не лежит константа $2$, 
а все функции этого основания сохраняют некоторое 
нетривиальное разбиение отличное от $\{0,1\}\{2\}$, 
то максимальный надкласс такого основания неявно неполон.
\end{lemma}

\begin{proof}
Из первой части условия можно получить, 
что данный замкнутый класс --- подкласс $T_{\{0,1\},0}$.
Из второй части получим, 
что все функции этого класса 
(а не только основание) сохраняют некоторое 
нетривиальное разбиение отличное от $\{0,1\}\{2\}$.

Действительно, без ограничения общности предположим, 
что все функции основания сохраняют разбиение $\{0,2\}\{1\}$, 
и, кроме того, в классе есть функция, не сохраняющее это разбиение.
Тогда по лемме \ref{U1by2const} из этой функции 
с помощью констант $0$ и $1$ можно получить одноместную функцию, 
не сохраняющую это разбиение, 
что противоречит нашему предположению.

Покажем, что $T_{\{0,1\},0}\cap U_{\{0,2\}\{1\}}=
\Sigma^{\{0,2\}\{1\}}_{T_0}$.

$T_{\{0,1\},0}\cap U_{\{0,2\}\{1\}}\subseteq
\Sigma^{\{0,2\}\{1\}}_{T_0}$ так как, во-первых, любая функция из 
$T_{\{0,1\},0}\cap U_{\{0,2\}\{1\}}$ сохраняет разбиение 
$\{0,2\}\{1\}$, а во-вторых, 
на наборах из $\{0,2\}^n$ принимает значения из 
множества $\{0,2\}$. 
Наборы такого вида суть нулевые наборы на всех блоках. 
То есть какой бы блок мы ни взяли, 
ему либо сопоставляется константа $2$, 
либо булева функция, 
которая на нулевом наборе принимает значение $0$.
А значит эта функция из $\Sigma^{\{0,2\}\{1\}}_{T_0}$.

С другой стороны, $T_{\{0,1\},0}\cap U_{\{0,2\}\{1\}}
\supseteq \Sigma^{\{0,2\}\{1\}}_{T_0}$, так как всякая функция из 
$\Sigma^{\{0,2\}\{1\}}_{T_0}$ по определению сохраняет разбиение 
$\{0,2\}\{1\}$, а на нулевых наборах всех своих блоков 
(то есть на наборах из $\{0,2\}^n$) она равна либо $0$, либо $2$.

Аналогично $T_{\{0,1\},0}\cap U_{\{1,2\}\{0\}}=
\Sigma^{\{1,2\}\{0\}}_{T_0}$.

Таким образом, класс, удовлетворяющий условию леммы, 
содержится в неявно неполном классе, а, следовательно, 
и сам неявно неполон.
\end{proof}

Воспользуемся доказанными леммами и покажем,
почему основания из групп 1 -- 5 не могут 
являться основаниями неявно предполных классов 
отличных от указанных в теореме~\ref{all2const}.

Основания 1.1 -- 1.13 --- по лемме~\ref{2const1} б).

Основания 2.1 -- 2.6 --- по лемме~\ref{2const1} а).

Основания 3.1 -- 3.8 --- по лемме~\ref{2const1} в).

Основания 4.1 -- 4.11 --- по леммам~\ref{2const32} и~\ref{2const3}.

Основания 5.1 -- 5.4 --- по лемме~\ref{2const4}.

Основание 6.1 --- по лемме~\ref{2const2}.

\subsection{Классы с тремя константами}
Так как различных оснований, 
содержащих 3 константы, очень много, разобьем 
этот случай на несколько подслучаев, 
в каждом из которых будем рассматривать основания, 
содержащиеся в одном из предполных по суперпозиции классов.

\subsubsection{Классы типа U}
\begin{theorem}\label{allU}
Существует ровно 15 неявно предполных классов, 
содержащих все константы, основания которых целиком содержатся в классах 
типа $U$:
$\Sigma^{\{0,1\}\{2\}}_K$, $\Sigma^{\{0,2\}\{1\}}_K$, 
$\Sigma^{\{1,2\}\{0\}}_K$, $\Sigma^{\{0,1\}\{2\}}_D$,
$\Sigma^{\{0,2\}\{1\}}_D$, $\Sigma^{\{1,2\}\{0\}}_D$,
$\Sigma^{\{0,1\}\{2\}}_L$, $\Sigma^{\{0,2\}\{1\}}_L$, 
$\Sigma^{\{1,2\}\{0\}}_L$, $W_0$, $W_1$, $W_2$, $F'_0$, $F'_1$, $F'_2$.
\end{theorem}

\begin{proof}
Множество одноместных функций класса $U_{\{0,1\}\{2\}}$ 
отличается от множества одноместных функций класса 
$T_{\{0,1\},0}$ только константой $2$ и парой функций 
$\szzo\szzi$. 
Таким образом, основания, которые нас интересуют получаются 
из оснований предыдущей теоремы 
добавлением константы $2$ и, возможно, 
функций 
$\szzo\szzi$.  
Кроме того, функцию 
$
\szzo$ можно добавить к основанию тогда и только тогда, 
когда оно содержит функцию $\sooz$, 
так как первая из этих функций порождает вторую.
Аналогично функцию 
$\szzi$ можно добавить к основанию тогда и только тогда, 
когда оно содержит функцию 
$\siiz$.

Леммы, доказанные по ходу предыдущей теоремы существенно помогают 
в рассмотрении оснований из нынешней. Только вместо леммы \ref{Tdizkon} будет 
использоваться алогичная лемма \ref{Udizkon}. Рассмотрим, 
почему получаемые основания не могут быть основаниями других неявно предполных
классов:

Основания, полученные из 1.1 -- 1.13 --- по лемме~\ref{2const1} б).

Основания, полученные из 2.1 -- 1.6 --- по лемме~\ref{2const1} а).

Основания, полученные из 3.1 -- 3.8 --- по лемме~\ref{2const1} в).

Основания, полученные из 4.1 -- 4.11 --- 
по леммам~\ref{2const32} и~\ref{2const3}.

Классы с основаниями, полученными из 5.1 -- 5.4 --- 
по лемме~\ref{Umax} сохраняют два разбиения и 
содержатся в соответствующих классах $W_a$.

Основания, полученные из 6.1 --- по лемме~\ref{2const2}.

Нерассмотренными остались только основания, получаемые из 6.2 и 6.3:
$$
\boxed
{
\begin{matrix}
0\\
0\\
0
\end{matrix}
}
\boxed
{
\begin{matrix}
0\\
1\\
2
\end{matrix}
}
\boxed
{
\begin{matrix}
1\\
1\\
1
\end{matrix}
}
\boxed
{
\begin{matrix}
0\\
0\\
2
\end{matrix}
}
\boxed
{
\begin{matrix}
1\\
1\\
2
\end{matrix}
}
\boxed
{
\begin{matrix}
2\\
2\\
2
\end{matrix}
};\quad
\boxed
{
\begin{matrix}
0\\
0\\
0
\end{matrix}
}
\boxed
{
\begin{matrix}
0\\
1\\
2
\end{matrix}
}
\boxed
{
\begin{matrix}
1\\
1\\
1
\end{matrix}
}
\boxed
{
\begin{matrix}
0\\
0\\
2
\end{matrix}
}
\boxed
{
\begin{matrix}
1\\
1\\
2
\end{matrix}
}
\boxed
{
\begin{matrix}
2\\
2\\
0
\end{matrix}
}
\boxed
{
\begin{matrix}
2\\
2\\
1
\end{matrix}
}
\boxed
{
\begin{matrix}
2\\
2\\
2
\end{matrix}
};\quad
\boxed
{
\begin{matrix}
0\\
0\\
0
\end{matrix}
}
\boxed
{
\begin{matrix}
0\\
1\\
2
\end{matrix}
}
\boxed
{
\begin{matrix}
1\\
1\\
1
\end{matrix}
}
\boxed
{
\begin{matrix}
0\\
0\\
2
\end{matrix}
}
\boxed
{
\begin{matrix}
1\\
0\\
2
\end{matrix}
}
\boxed
{
\begin{matrix}
1\\
1\\
2
\end{matrix}
}
\boxed
{
\begin{matrix}
2\\
2\\
2
\end{matrix}
};\quad
\boxed
{
\begin{matrix}
0\\
0\\
0
\end{matrix}
}
\boxed
{
\begin{matrix}
0\\
1\\
2
\end{matrix}
}
\boxed
{
\begin{matrix}
1\\
1\\
1
\end{matrix}
}
\boxed
{
\begin{matrix}
0\\
0\\
2
\end{matrix}
}
\boxed
{
\begin{matrix}
1\\
0\\
2
\end{matrix}
}
\boxed
{
\begin{matrix}
1\\
1\\
2
\end{matrix}
}
\boxed
{
\begin{matrix}
2\\
2\\
0
\end{matrix}
}
\boxed
{
\begin{matrix}
2\\
2\\
1
\end{matrix}
}
\boxed
{
\begin{matrix}
2\\
2\\
2
\end{matrix}
}
$$

Ко второму и четвертому из этих оснований можно применить 
утверждение \ref{2const33}.

Максимальный надкласс третьего основания в точности совпадает с 
неявно предполным классом $F_2$.
Следовательно, класс с таким основанием либо совпадает с $F_2$, 
либо содержится в нем и не является неявно предполным.

Докажем, что максимальный надкласс первого основания 
содержится в классе $F_2$.

Предположим противное. Пусть существует функция, которая не сохраняет матрицу 
$
A_3=
\begin{pmatrix}
0 & 1 & 0 & 0 & 1 & 1 & 2\\
0 & 1 & 0 & 1 & 0 & 0 & 2\\
0 & 1 & 2 & 2 & 2 & 2 & 2
\end{pmatrix}
$, но при этом сохраняет 
$
\hat{A_3}=
\begin{pmatrix}
0 & 1 & 0 & 0 & 1 & 2\\
0 & 1 & 0 & 1 & 1 & 2\\
0 & 1 & 2 & 2 & 2 & 2
\end{pmatrix}
$.

Тогда при подстановке в функцию столбцов из $A_3$ в определенном порядке
результирующий столбец 
$
\begin{pmatrix}
a\\
b\\
c
\end{pmatrix}
$ не будет принадлежать $A_3$. Отождествим в функции переменные, 
соответствующие одинаковым столбцам, получим функцию 
$f(x_1,\dots,x_7)$ с такими же свойствами.

Построим функцию 
$
g(x,y)=f(
0,1,
\boxed
{
\begin{matrix}
0\\
0\\
2
\end{matrix}
}(x),
x,
y,
\boxed
{
\begin{matrix}
1\\
1\\
2
\end{matrix}
}(x),
2)=
\boxed
{
\begin{matrix}
* & a & *\\
b & * & *\\
* & * & c
\end{matrix}
}
$. Покажем, что, каким бы ни был столбец 
$
\boxed
{
\begin{matrix}
a\\
b\\
c
\end{matrix}
}\notin A_3
$, функция $g(x,y)$ не сохраняет $\hat{A_3}$.

Если $a=2, b\neq 2$ или наоборот, то $g(x,y)$ не сохраняет 
разбиение $\{0,1\}\{2\}$, что по лемме \ref{Umax} должны делать 
все функции класса с таким основанием.

Если $a=b=2, c\neq 2$, то $g(0,0)=g(1,1)=2$ (иначе $g$ не сохраняет разбиение).
Тогда $g(x,x)$ имеет вид 
$
\boxed
{
\begin{matrix}
2\\
2\\
c
\end{matrix}
}, c\in \{0,1\}
$. Но таких функций в основании нет.

Если $a,b,c\in \{0,1\}$ и среди них есть различные. Тогда одна из 
одноместных функций 
$
g(x,
\boxed
{
\begin{matrix}
1\\
1\\
2
\end{matrix}
}(x)),
g(
\boxed
{
\begin{matrix}
1\\
1\\
2
\end{matrix}
}(y),
y)
$ не лежит в основании. 

Таким образом, не существует функции, которая могла бы принадлежать 
классу с основанием $\hat{A_3}$, но не принадлежать классу $F_2$.

\medskip
Все основания рассмотрены, теорема доказана.
\end{proof}

\subsubsection{Классы типа M}
\begin{theorem}\label{allM}
Существует ровно 6 неявно предполных классов в $P_3$, 
содержащих все константы, основания которых целиком содержатся 
хотя бы в одном из классов типа $M$, 
но не содержатся ни в одном из классов 
типа $U$:$KM_1$, $KM_2$, $KM_3$, $DM_1$, $DM_2$, $DM_3$.
\end{theorem}

\begin{proof}
Можно показать, что основание явно замкнутого класса, 
содержащего все 3 константы, 
содержащееся в классе $M_1$, 
но не содержащееся целиком ни в одном из классов типа $U$, 
совпадает с одним из следующих трех оснований 
(или двойственно ему):

$$
\boxed
{
\begin{matrix}
0\\
0\\
0
\end{matrix}
}
\boxed
{
\begin{matrix}
0\\
0\\
1
\end{matrix}
}
\boxed
{
\begin{matrix}
0\\
1\\
1
\end{matrix}
}
\boxed
{
\begin{matrix}
0\\
1\\
2
\end{matrix}
}
\boxed
{
\begin{matrix}
1\\
1\\
1
\end{matrix}
}
\boxed
{
\begin{matrix}
1\\
1\\
2
\end{matrix}
}
\boxed
{
\begin{matrix}
1\\
2\\
2
\end{matrix}
}
\boxed
{
\begin{matrix}
2\\
2\\
2
\end{matrix}
};\;
\boxed
{
\begin{matrix}
0\\
0\\
0
\end{matrix}
}
\boxed
{
\begin{matrix}
0\\
0\\
2
\end{matrix}
}
\boxed
{
\begin{matrix}
0\\
1\\
2
\end{matrix}
}
\boxed
{
\begin{matrix}
0\\
2\\
2
\end{matrix}
}
\boxed
{
\begin{matrix}
1\\
1\\
1
\end{matrix}
}
\boxed
{
\begin{matrix}
1\\
1\\
2
\end{matrix}
}
\boxed
{
\begin{matrix}
1\\
2\\
2
\end{matrix}
}
\boxed
{
\begin{matrix}
2\\
2\\
2
\end{matrix}
};\;
\boxed
{
\begin{matrix}
0\\
0\\
0
\end{matrix}
}
\boxed
{
\begin{matrix}
0\\
0\\
1
\end{matrix}
}
\boxed
{
\begin{matrix}
0\\
0\\
2
\end{matrix}
}
\boxed
{
\begin{matrix}
0\\
1\\
1
\end{matrix}
}
\boxed
{
\begin{matrix}
0\\
1\\
2
\end{matrix}
}
\boxed
{
\begin{matrix}
0\\
2\\
2
\end{matrix}
}
\boxed
{
\begin{matrix}
1\\
1\\
1
\end{matrix}
}
\boxed
{
\begin{matrix}
1\\
1\\
2
\end{matrix}
}
\boxed
{
\begin{matrix}
1\\
2\\
2
\end{matrix}
}
\boxed
{
\begin{matrix}
2\\
2\\
2
\end{matrix}
}.
$$

Покажем, 
что неявно предполными с такими основаниями могут быть 
только классы, 
указанные в условии теоремы. Будем искать классы, 
которые целиком не содержатся в неявно предполных классах 
$KM_1, DM_1$ и $R'_2$ и, соответственно, содержат функции 
$f_K(\tilde{x})\notin KM_1, f_D(\tilde{x})\notin DM_1$ и 
$f_R(\tilde{x})\notin R'_2$.

Нам потребуется вспомогательная
\begin{lemma}
Если основание явно замкнутого класса целиком содержится в 
классе типа $M$, а также содержит все 3 константы, то и сам класс 
целиком содержится в классе типа $M$.
\end{lemma}
\begin{proof}
Докажем это утверждение для классов, монотонных относительно порядка $0<1<2$.
Для остальных монотонных классов это будет 
следовать из принципа двойственности. 

Предположим, что в классе содержится немонотонная функция $g(\tilde{x})$. 
Тогда существует такая пара наборов $\tilde{\alpha}\prec \tilde{\beta}$, что 
$g(\tilde{\alpha})>g(\tilde{\beta})$. 
Причем обязательно можно найти два соседних таких набора. 
Без ограничения общности можно считать, что 
$$
\alpha=(0,\dots,0,1,\dots,1,2,\dots,2,a),
$$
$$
\beta=(0,\dots,0,1,\dots,1,2,\dots,2,b),
$$
где $a<b$. 
Тогда рассмотрим $g(0,\dots,0,1,\dots,1,2,\dots,2,x)=g'(x)$. 
Такая одноместная функция должна лежать в нашем классе, 
так все функции, входящие в суперпозицию, также в него входят. 
Но $g'(x)$ немонотонна по построению, 
а значит не может лежать в нашем 
основании. Противоречие.
\end{proof}

\begin{lemma}\label{M1base}
Явно замкнутый класс с основанием I не может быть неявно предполным.
\end{lemma}
\begin{proof}
Рассмотрим функцию $f_K(\tilde{x})$. 
Пусть она не сохраняет матрицу
$
A_4=
\begin{pmatrix}
0 & 0 & 0 & 1 & 1 & 1 & 2 & 2 & 2\\
0 & 1 & 2 & 0 & 1 & 2 & 0 & 1 & 2\\
0 & 0 & 0 & 0 & 1 & 1 & 0 & 1 & 2
\end{pmatrix}
$, а значит можно без ограничения общности считать, что зависит от 
9 переменных. Пусть
\begin{gather*}
f_K(0,0,0,1,1,1,2,2,2)=a,\\
f_K(0,1,2,0,1,2,0,1,2)=b,\\
f_K(0,0,0,0,1,1,0,1,2)=c,
\end{gather*}
тогда столбец 
$
\begin{pmatrix}
a\\
b\\
c
\end{pmatrix}
\notin A_4
$.

Можно построить функцию 
$$
g_K(x,y)=f_K(0,
\boxed
{
\begin{matrix}
0\\
0\\
1
\end{matrix}
}(x),
x,
\boxed
{
\begin{matrix}
0\\
0\\
1
\end{matrix}
}(y),
1,
\boxed
{
\begin{matrix}
1\\
1\\
2
\end{matrix}
}(x),
y,
\boxed
{
\begin{matrix}
1\\
1\\
2
\end{matrix}
}(y),
2)=
\boxed
{
\begin{matrix}
c & * & a\\
* & * & *\\
b & * & *
\end{matrix}
}.
$$ 
Отсюда видно, что $c\le min(a,b)$, так как иначе подстановкой констант 
мы бы получили немонотонную функцию. Кроме того $c\neq min(a,b)$, 
так как все такие столбцы принадлежат матрице $A_4$. 
Таким образом, столбец
$
\begin{pmatrix}
a\\
b\\
c
\end{pmatrix}
$
совпадает с одним из столбцов
$
\left\lbrace
\begin{pmatrix}
1\\
1\\
0
\end{pmatrix},
\begin{pmatrix}
1\\
2\\
0
\end{pmatrix},
\begin{pmatrix}
2\\
1\\
0
\end{pmatrix},
\begin{pmatrix}
2\\
2\\
0
\end{pmatrix},
\begin{pmatrix}
2\\
2\\
1
\end{pmatrix}
\right\rbrace
$.

Если $
\begin{pmatrix}
a\\
b\\
c
\end{pmatrix}
\in 
\left\lbrace
\begin{pmatrix}
1\\
1\\
0
\end{pmatrix},
\begin{pmatrix}
1\\
2\\
0
\end{pmatrix},
\begin{pmatrix}
2\\
1\\
0
\end{pmatrix},
\begin{pmatrix}
2\\
2\\
0
\end{pmatrix}
\right\rbrace
$, то 
$
\boxed
{
\begin{matrix}
0\\
1\\
1
\end{matrix}
}
(g_K(x,y))
$ имеет вид 
$
\boxed
{
\begin{matrix}
0   & a_k & 1\\
b_k & c_k & 1\\
1   & 1   & 1
\end{matrix}
}
$.

Если $
\begin{pmatrix}
a\\
b\\
c
\end{pmatrix}
=
\begin{pmatrix}
2\\
2\\
1
\end{pmatrix}
$, то 
$
\boxed
{
\begin{matrix}
0\\
0\\
1
\end{matrix}
}
(g_K(x,y))
$ имеет вид 
$
\boxed
{
\begin{matrix}
0   & a_k & 1\\
b_k & c_k & 1\\
1   & 1   & 1
\end{matrix}
}
=h_K(x,y)
$. 

Причем
$
\begin{pmatrix}
a_k\\
b_k\\
c_k
\end{pmatrix}
\in 
\left\lbrace
\begin{pmatrix}
0\\
0\\
0
\end{pmatrix},
\begin{pmatrix}
0\\
0\\
1
\end{pmatrix},
\begin{pmatrix}
0\\
1\\
1
\end{pmatrix},
\begin{pmatrix}
1\\
0\\
1
\end{pmatrix},
\begin{pmatrix}
1\\
1\\
1
\end{pmatrix}
\right\rbrace
$.

Покажем, что каким бы ни был набор 
$
(a_k,b_k,c_k)
$, из функции вида 
$
\boxed
{
\begin{matrix}
0   & a_k & 1\\
b_k & c_k & 1\\
1   & 1   & 1
\end{matrix}
}
$ и функций из основания I посредством суперпозиций 
всегда можно получить функцию 
$
\boxed
{
\begin{matrix}
0 & 1 & 1\\
1 & 1 & 1\\
1 & 1 & 1
\end{matrix}
}
$.

Если
$
\begin{pmatrix}
a_k\\
b_k\\
c_k
\end{pmatrix}
=
\begin{pmatrix}
0\\
0\\
0
\end{pmatrix}
$, то 
$
h_K(
\boxed
{
\begin{matrix}
1\\
1\\
2
\end{matrix}
}(x),
\boxed
{
\begin{matrix}
1\\
1\\
2
\end{matrix}
}(y)
)=
\boxed
{
\begin{matrix}
0 & 1 & 1\\
1 & 1 & 1\\
1 & 1 & 1
\end{matrix}
}
$.

Если
$
\begin{pmatrix}
a_k\\
b_k\\
c_k
\end{pmatrix}
=
\begin{pmatrix}
0\\
0\\
1
\end{pmatrix}
$, то 
$
h_K(
\boxed
{
\begin{matrix}
1\\
1\\
2
\end{matrix}
}(x),
y
)=
\boxed
{
\begin{matrix}
0 & 1 & 1\\
1 & 1 & 1\\
1 & 1 & 1
\end{matrix}
}
$.

Если
$
\begin{pmatrix}
a_k\\
b_k\\
c_k
\end{pmatrix}
=
\begin{pmatrix}
1\\
0\\
1
\end{pmatrix}
$, то 
$
h_K(
\boxed
{
\begin{matrix}
1\\
1\\
2
\end{matrix}
}(x),
y
)=
\boxed
{
\begin{matrix}
0 & 1 & 1\\
1 & 1 & 1\\
1 & 1 & 1
\end{matrix}
}
$.

Если
$
\begin{pmatrix}
a_k\\
b_k\\
c_k
\end{pmatrix}
=
\begin{pmatrix}
0\\
1\\
1
\end{pmatrix}
$, то 
$
h_K(x,
\boxed
{
\begin{matrix}
1\\
1\\
2
\end{matrix}
}(y)
)=
\boxed
{
\begin{matrix}
0 & 1 & 1\\
1 & 1 & 1\\
1 & 1 & 1
\end{matrix}
}
$.

Если
$
\begin{pmatrix}
a_k\\
b_k\\
c_k
\end{pmatrix}
=
\begin{pmatrix}
1\\
1\\
1
\end{pmatrix}
$, то 
$
h_K(x,y)=
\boxed
{
\begin{matrix}
0 & 1 & 1\\
1 & 1 & 1\\
1 & 1 & 1
\end{matrix}
}
$.

Аналогичным образом рассмотрим функцию $f_D$. 
Она не сохраняет матрицу
$
A'_4=
\begin{pmatrix}
0 & 0 & 0 & 1 & 1 & 1 & 2 & 2 & 2\\
0 & 1 & 2 & 0 & 1 & 2 & 0 & 1 & 2\\
0 & 1 & 2 & 1 & 1 & 2 & 2 & 2 & 2
\end{pmatrix}
$, а значит можно без ограничения общности считать, что зависит от 
9 переменных. Пусть
\begin{gather*}
f_D(0,0,0,1,1,1,2,2,2)=a',\\
f_D(0,1,2,0,1,2,0,1,2)=b',\\
f_D(0,1,2,1,1,2,2,2,2)=c',
\end{gather*}
тогда столбец 
$
\begin{pmatrix}
a'\\
b'\\
c'
\end{pmatrix}
\notin A'_4
$.

Можно построить функцию 
$$
g_D(x,y)=f_D(0,
\boxed
{
\begin{matrix}
0\\
0\\
1
\end{matrix}
}(x),
x,
\boxed
{
\begin{matrix}
0\\
0\\
1
\end{matrix}
}(y),
1,
\boxed
{
\begin{matrix}
1\\
1\\
2
\end{matrix}
}(x),
y,
\boxed
{
\begin{matrix}
1\\
1\\
2
\end{matrix}
}(y),
2)=
\boxed
{
\begin{matrix}
*  & * & a'\\
*  & * &  *\\
b' & * & c'
\end{matrix}
}.
$$ 
Отсюда видно, что $c'\ge max(a',b')$, так как иначе подстановкой констант 
мы бы получили немонотонную функцию. Кроме того $c'\neq max(a',b')$, 
так как все такие столбцы принадлежат матрице $A'_4$. 
Таким образом, столбец
$
\begin{pmatrix}
a'\\
b'\\
c'
\end{pmatrix}
$
совпадает с одним из столбцов
$
\left\lbrace
\begin{pmatrix}
0\\
0\\
1
\end{pmatrix},
\begin{pmatrix}
0\\
0\\
2
\end{pmatrix},
\begin{pmatrix}
0\\
1\\
2
\end{pmatrix},
\begin{pmatrix}
1\\
0\\
2
\end{pmatrix},
\begin{pmatrix}
1\\
1\\
2
\end{pmatrix}
\right\rbrace
$.

Если $
\begin{pmatrix}
a'\\
b'\\
c'
\end{pmatrix}
\in 
\left\lbrace
\begin{pmatrix}
0\\
0\\
2
\end{pmatrix},
\begin{pmatrix}
0\\
1\\
2
\end{pmatrix},
\begin{pmatrix}
1\\
0\\
2
\end{pmatrix},
\begin{pmatrix}
1\\
1\\
2
\end{pmatrix}
\right\rbrace
$, то 
$
\boxed
{
\begin{matrix}
0\\
0\\
1
\end{matrix}
}
(g_D(x,y))
$ имеет вид 
$
\boxed
{
\begin{matrix}
0 & 0   & 0\\
0 & a_d & b_d\\
0 & c_d & 1
\end{matrix}
}
$.

Если $
\begin{pmatrix}
a'\\
b'\\
c'
\end{pmatrix}
=
\begin{pmatrix}
0\\
0\\
1
\end{pmatrix}
$, то 
$
g_D(x,y)
$ имеет вид 
$
\boxed
{
\begin{matrix}
0 & 0   & 0\\
0 & a_d & b_d\\
0 & c_d & 1 
\end{matrix}
}
=h_D(x,y)
$. 

Причем
$
\begin{pmatrix}
a_d\\
b_d\\
c_d
\end{pmatrix}
\in 
\left\lbrace
\begin{pmatrix}
0\\
0\\
0
\end{pmatrix},
\begin{pmatrix}
0\\
0\\
1
\end{pmatrix},
\begin{pmatrix}
0\\
1\\
0
\end{pmatrix},
\begin{pmatrix}
0\\
1\\
1
\end{pmatrix},
\begin{pmatrix}
1\\
1\\
1
\end{pmatrix}
\right\rbrace
$.

Покажем, что каким бы ни был набор 
$
(a_d,b_d,c_d)
$, из функции вида 
$
\boxed
{
\begin{matrix}
0 & 0   & 0\\
0 & a_d & b_d\\
0 & c_d & 1
\end{matrix}
}
$ и функций из основания I посредством суперпозиций 
всегда можно получить функцию 
$
\boxed
{
\begin{matrix}
0 & 0 & 0\\
0 & 1 & 1\\
0 & 1 & 1
\end{matrix}
}
$.

Если
$
\begin{pmatrix}
a_d\\
b_d\\
c_d
\end{pmatrix}
=
\begin{pmatrix}
0\\
0\\
0
\end{pmatrix}
$, то 
$
h_D(
\boxed
{
\begin{matrix}
1\\
1\\
2
\end{matrix}
}(x),
\boxed
{
\begin{matrix}
1\\
1\\
2
\end{matrix}
}(y)
)=
\boxed
{
\begin{matrix}
0 & 0 & 0\\
0 & 1 & 1\\
0 & 1 & 1
\end{matrix}
}
$.

Если
$
\begin{pmatrix}
a_d\\
b_d\\
c_d
\end{pmatrix}
=
\begin{pmatrix}
0\\
0\\
1
\end{pmatrix}
$, то 
$
h_D(
\boxed
{
\begin{matrix}
1\\
1\\
2
\end{matrix}
}(x),
y
)=
\boxed
{
\begin{matrix}
0 & 0 & 0\\
0 & 1 & 1\\
0 & 1 & 1
\end{matrix}
}
$.

Если
$
\begin{pmatrix}
a_d\\
b_d\\
c_d
\end{pmatrix}
=
\begin{pmatrix}
0\\
1\\
1
\end{pmatrix}
$, то 
$
h_D(
\boxed
{
\begin{matrix}
1\\
1\\
2
\end{matrix}
}(x),
y
)=
\boxed
{
\begin{matrix}
0 & 0 & 0\\
0 & 1 & 1\\
0 & 1 & 1
\end{matrix}
}
$.

Если
$
\begin{pmatrix}
a_d\\
b_d\\
c_d
\end{pmatrix}
=
\begin{pmatrix}
0\\
1\\
0
\end{pmatrix}
$, то 
$
h_D(x,
\boxed
{
\begin{matrix}
1\\
1\\
2
\end{matrix}
}(y)
)=
\boxed
{
\begin{matrix}
0 & 0 & 0\\
0 & 1 & 1\\
0 & 1 & 1
\end{matrix}
}
$.

Если
$
\begin{pmatrix}
a_d\\
b_d\\
c_d
\end{pmatrix}
=
\begin{pmatrix}
1\\
1\\
1
\end{pmatrix}
$, то 
$
h_D(x,y)=
\boxed
{
\begin{matrix}
0 & 0 & 0\\
0 & 1 & 1\\
0 & 1 & 1
\end{matrix}
}
$.

Таким образом, в искомом классе должны лежать функции
$
\boxed
{
\begin{matrix}
0 & 0 & 0\\
0 & 1 & 1\\
0 & 1 & 1
\end{matrix}
}
\boxed
{
\begin{matrix}
0 & 1 & 1\\
1 & 1 & 1\\
1 & 1 & 1
\end{matrix}
}
\boxed
{
\begin{matrix}
0\\
0\\
1
\end{matrix}
}
\boxed
{
\begin{matrix}
1\\
1\\
1
\end{matrix}
}
$, которые порождают  неявно полный класс $S_2$.
Следовательно, и сам класс оказывается неявно полным.
\end{proof}

\begin{lemma}
Явно замкнутый класс с основанием II 
не может быть неявно предполным.
\end{lemma}
\begin{proof}
Так же, как и в предыдущей лемме рассмотрим функцию
$f_K(\tilde{x})$,
не сохраняющую $A_4$. 
Из нее с помощью функций из основания построим 
$
g_K(x,y)=f_K(
0,x,
\boxed
{
\begin{matrix}
0\\
2\\
2
\end{matrix}
}(x),
y,1,
\boxed
{
\begin{matrix}
1\\
2\\
2
\end{matrix}
}(x),
\boxed
{
\begin{matrix}
0\\
2\\
2
\end{matrix}
}(y),
\boxed
{
\begin{matrix}
1\\
2\\
2
\end{matrix}
}(y),
2)
=
\boxed
{
\begin{matrix}
c & a & *\\
b & * & *\\
* & * & *
\end{matrix}
}
$.

Столбец 
$
\begin{pmatrix}
a\\
b\\
c
\end{pmatrix}
$ не принадлежит матрице $A_4$. 
Повторяя рассуждения из предыдущей леммы, 
снова получим, что
$
\begin{pmatrix}
a\\
b\\
c
\end{pmatrix}
\in 
\left\lbrace
\begin{pmatrix}
1\\
1\\
0
\end{pmatrix},
\begin{pmatrix}
1\\
2\\
0
\end{pmatrix},
\begin{pmatrix}
2\\
1\\
0
\end{pmatrix},
\begin{pmatrix}
2\\
2\\
0
\end{pmatrix},
\begin{pmatrix}
2\\
2\\
1
\end{pmatrix}
\right\rbrace
$.

Если 
$
\begin{pmatrix}
a\\
b\\
c
\end{pmatrix}
=
\begin{pmatrix}
2\\
2\\
1
\end{pmatrix}
$, то
$
\boxed
{
\begin{matrix}
0\\
0\\
2
\end{matrix}
}
(g_K(x,y))=
\boxed
{
\begin{matrix}
0 & 2 & 2\\
2 & 2 & 2\\
2 & 2 & 2
\end{matrix}
}
$.

Если же 
$
\begin{pmatrix}
a\\
b\\
c
\end{pmatrix}
\neq
\begin{pmatrix}
2\\
2\\
1
\end{pmatrix}
$, то
$
\boxed
{
\begin{matrix}
0\\
2\\
2
\end{matrix}
}
(g_K(x,y))=
\boxed
{
\begin{matrix}
0 & 2 & 2\\
2 & 2 & 2\\
2 & 2 & 2
\end{matrix}
}
$.

Таким образом, какой бы ни была функция $f_K$, из нее с помощью
функций основания всегда можно получить функцию 
$
\boxed
{
\begin{matrix}
0 & 2 & 2\\
2 & 2 & 2\\
2 & 2 & 2
\end{matrix}
}
$.
\smallskip

Можно заметить, что все функции основания лежат в классе $R'_2$, 
поэтому, раз мы ищем класс отличный от описанных, в нем должна быть 
функция $f_R\notin R'_2$.

Такая функция должна не сохранять матрицу
$
A_1=
\begin{pmatrix}
0 & 1 & 0 & 1 & 2 & 2 & 2\\
0 & 1 & 2 & 2 & 0 & 1 & 2\\
0 & 1 & 2 & 2 & 2 & 2 & 2
\end{pmatrix}
$, а значит без ограничения общности можно считать, что $f_R$ 
зависит от 7 переменных. Пусть
\begin{gather*}
f_R(0,1,0,1,2,2,2)=a',\\
f_R(0,1,2,2,0,1,2)=b',\\
f_R(0,1,2,2,2,2,2)=c'.
\end{gather*}

Из $f_R$ построим функцию 
$
g_R(x,y)=
f_R(0,1,
\boxed
{
\begin{matrix}
0\\
0\\
2
\end{matrix}
}(x),
x,
\boxed
{
\begin{matrix}
0\\
0\\
2
\end{matrix}
}(y),
y,2)=
\boxed
{
\begin{matrix}
* &  * & * \\
* &  * & a'\\
* & b' & c'
\end{matrix}
}
$.

При этом столбец
$
\begin{pmatrix}
a'\\
b'\\
c'
\end{pmatrix}
$
должен, с одной стороны, удовлетворять условию $c\ge max(a,b)$, 
так как иначе у нас бы получилась немонотонная функция, а с другой, 
не принадлежать $A_1$. Отсюда получим, что
$$
\begin{pmatrix}
a'\\
b'\\
c'
\end{pmatrix}
\in 
\left\lbrace
\begin{pmatrix}
0\\
0\\
1
\end{pmatrix},
\begin{pmatrix}
0\\
0\\
2
\end{pmatrix},
\begin{pmatrix}
0\\
1\\
1
\end{pmatrix},
\begin{pmatrix}
0\\
1\\
2
\end{pmatrix},
\begin{pmatrix}
1\\
0\\
1
\end{pmatrix},
\begin{pmatrix}
1\\
0\\
2
\end{pmatrix},
\begin{pmatrix}
1\\
1\\
2
\end{pmatrix}
\right\rbrace
$$

Если 
$
\begin{pmatrix}
a'\\
b'\\
c'
\end{pmatrix}
\in 
\left\lbrace
\begin{pmatrix}
0\\
1\\
1
\end{pmatrix},
\begin{pmatrix}
1\\
0\\
1
\end{pmatrix}
\right\rbrace
$, то подстановкой констант можно получить одноместную функцию 
$
\boxed
{
\begin{matrix}
0\\
0\\
1
\end{matrix}
}
$, которой нет в основании.

Если
$
\begin{pmatrix}
a'\\
b'\\
c'
\end{pmatrix}
\in 
\left\lbrace
\begin{pmatrix}
0\\
0\\
2
\end{pmatrix},
\begin{pmatrix}
0\\
1\\
2
\end{pmatrix},
\begin{pmatrix}
1\\
0\\
2
\end{pmatrix},
\begin{pmatrix}
1\\
1\\
2
\end{pmatrix}
\right\rbrace
$, то 
$
\boxed
{
\begin{matrix}
0\\
0\\
2
\end{matrix}
}
(g_R(x,y))=
\boxed
{
\begin{matrix}
0 & 0 & 0\\
0 & 0 & 0\\
0 & 0 & 2
\end{matrix}
}
$.

Если 
$
\begin{pmatrix}
a'\\
b'\\
c'
\end{pmatrix}
=
\begin{pmatrix}
0\\
0\\
1
\end{pmatrix}
$, то 
$\boxed
{
\begin{matrix}
0\\
2\\
2
\end{matrix}
}
(g_R(x,y))=
\boxed
{
\begin{matrix}
0 & 0 & 0\\
0 & 0 & 0\\
0 & 0 & 2
\end{matrix}
}
$.

Таким образом, какой бы ни была функция $f_R$, из нее с помощью 
функций основания всегда можно получить функцию
$
\boxed
{
\begin{matrix}
0 & 0 & 0\\
0 & 0 & 0\\
0 & 0 & 2
\end{matrix}
}
$. А значит в нашем классе найдутся функции
$
\boxed
{
\begin{matrix}
0 & 0 & 0\\
0 & 0 & 0\\
0 & 0 & 2
\end{matrix}
}
\boxed
{
\begin{matrix}
0 & 2 & 2\\
2 & 2 & 2\\
2 & 2 & 2
\end{matrix}
}
\boxed
{
\begin{matrix}
0\\
0\\
0
\end{matrix}
}
\boxed
{
\begin{matrix}
2\\
2\\
2
\end{matrix}
}
$, которые порождают неявно полный класс, 
двойственный $S_1$ относительно подстановки $(02)$.
\end{proof}

\begin{lemma}
Явно замкнутый класс с основанием III, отличный от классов $KM_1$ и $DM_1$ 
не может быть неявно предполным.
\end{lemma}
\begin{proof}
Поскольку основание III целиком содержит в себе основание I, доказательство 
леммы \ref{M1base} дословно проходит и для доказательства этого случая.
\end{proof}

Таким образом, все возможные основания рассмотрены. Теорема доказана.
\end{proof}

\subsubsection{Классы типа T1}
\begin{theorem}\label{allT1}
Существует ровно 6 неявно предполных классов в $P_3$, 
содержащих все константы, 
основания которых целиком содержатся хотя бы в одном 
из классов типа $T_{\{a\},1}$, 
но не содержатся ни в одном из классов 
типа $U$ и $M$: $R'_0$, $R'_1$, $R'_2$, $Q'_0$, $Q'_1$, $Q'_2$.
\end{theorem}

\begin{proof}
Можно показать, что основание явно замкнутого класса, 
содержащего все 3 константы, 
содержащееся в классе $T_{\{2\},1}$, 
но не содержащееся целиком ни в одном из классов типа $U$ и $M$, 
получается из следующих десяти оснований 
(и двойственных им) добавлением констант $0, 1, 2$, 
а также функции $x$:
\begin{gather*}
1.1-1.4\\
\sozz\sziz\sizz\szoz;\quad
\sozz\sziz\sizz\szoz\sioz;\quad
\sozz\sziz\sizz\szii\szoz\sozo\szoo\sizi;\quad
\sozz\sziz\sizz\szii\szoz\sozo\szoo\sizi\sioz.
\\
2.1-2.2\\
\sooz\sozz\siiz\szoo\szii\szzo\szzi;\quad
\sooz\sozz\siiz\szoo\szii\szzo\szzi\sioz
\end{gather*}
\begin{gather*}
3.1-3.4\\
\sooz\sozz\siiz\sziz\sizz\szoz;\quad
\sooz\sozz\siiz\sziz\sizz\szoz\sioz;\quad
\\
\sooz\sozo\siiz\sziz\sizz\szoz\sozo\sizi\szoo\szii\szzo\szzi;\quad
\sooz\sozz\siiz\sziz\sizz\szoz\sozo\sizi\szoo\szii\szzo\szzi\sioz
\end{gather*}

\begin{lemma}\label{kvazi}
Пусть явно замкнутый класс не содержится в классе $\mathfrak{N}$ 
квазилинейных функций, 
и при этом в его основании содержится функция $\phi$, 
принимающая на наборах $0$ и $2$ значения $1$ и $2$ 
(необязательно в таком порядке), 
а также все константы. 
Тогда для его неявной полноты достаточно, 
чтобы он содержал функцию вида
$
\boxed
{
\begin{matrix}
2\\
*\\
0
\end{matrix}
}
$, а также любые две функции из следующих трех:
$
\boxed
{
\begin{matrix}
0\\
0\\
2
\end{matrix}
}
\boxed
{
\begin{matrix}
0\\
2\\
2
\end{matrix}
}
\boxed
{
\begin{matrix}
2\\
0\\
2
\end{matrix}
}
$.

\end{lemma}
\begin{proof}
Пусть функция $f_n(\tilde{x})$ выделяет вершину некоторого квадрата. 
По определению квадрата 
у его наборов все компоненты кроме двух совпадают.
Подставим в $f_n$ на место этих компонент константы 
и получим функцию $g_n(x,y)$ из нашего класса, 
которая также не принадлежит $\mathfrak{N}$.

Построим теперь функцию двух переменных, 
которая выделяет вершину квадрата 
$(0,0);(0,2);(2,0);(2,2)$.

\begin{statement}
Если функция от двух переменных в $P_3$ выделяет вершину квадрата,
который не содержит вершину $(2,2)$, то она также выделяет вершину 
некоторого квадрата, содержащего вершину $(2,2)$.
\end{statement}
\begin{proof} 
Предположим, что исходная функция принимала всего два значения --- $0$ и $1$. 
Очевидно, что если утверждение верно для функции с таким ограничением, 
то оно верно и в общем случае. Тогда то, что функция выделяет 
вершину некоторого квадрата, эквивалентно тому, что сумма значений функции 
на наборах из этого квадрата нечетна. 

Пусть наша функция имеет вид 
$
\boxed
{
\begin{matrix}
a_{00} & a_{01} & a_{02}\\
a_{10} & a_{11} & a_{12}\\
a_{20} & a_{21} & a_{22}
\end{matrix}
}
$ и не выделяет квадратов с вершиной $(2,2)$. Тогда
\begin{gather*}
a_{00}+a_{02}+a_{20}+a_{22} \equiv 0 (mod 2);\\
a_{01}+a_{02}+a_{21}+a_{22} \equiv 0 (mod 2);\\
a_{10}+a_{12}+a_{20}+a_{22} \equiv 0 (mod 2);\\
a_{11}+a_{12}+a_{21}+a_{22} \equiv 0 (mod 2).
\end{gather*}
Отсюда
\begin{gather*}
a_{00}+a_{01}+2a_{02}+a_{20}+a_{21}+2a_{22}\equiv 0 (mod 2);\\
a_{00}+a_{01}+a_{20}+a_{21}\equiv 0 (mod 2),
\end{gather*}
а значит эта функция не выделяет вершину квадрата 
$(0,0);(0,1);(2,0);(2,1)$. Аналогично, складывая другие пары уравнений 
(или все четыре), получим, что функция не выделяет ни одного квадрата.
Противоречие.
\end{proof}

С учетом доказанного утверждения можно считать, что правый нижний угол 
нашего квадрата --- набор $(2,2)$. Пусть левый верхний --- набор $(b_1, b_2)$.
 
Рассмотрим функцию 
$h(x,y)=g_n(\psi_i (x),\psi_i(y))$, где 
$$
\psi_i(z)=
\begin{cases}
z, &\text{если $b_i=0$;}\\
\phi(z), &\text{если $b_i=1$.}
\end{cases}
$$
Она выделяет вершину квадрата, содержащего наборы $(0,0)$ и $(2,2)$. 

Теперь из $h(x,y)$ и имеющихся одноместных функций выразим пару функций 
$m(x,y)$ вида 
$
\boxed
{
\begin{matrix}
0 & * & 0\\
* & * & *\\
0 & * & 2
\end{matrix}
}
$ и $M(x,y)$ вида 
$
\boxed
{
\begin{matrix}
0 & * & 2\\
* & * & *\\
2 & * & 2
\end{matrix}
}
$.

Предположим, что у нас есть все три функции 
$
\boxed
{
\begin{matrix}
0\\
0\\
2
\end{matrix}
}
\boxed
{
\begin{matrix}
0\\
2\\
2
\end{matrix}
}
\boxed
{
\begin{matrix}
2\\
0\\
2
\end{matrix}
}
$. Тогда, если значение в вершине, которую выделяет $h(x,y)$ равно $0$, 
рассмотрим 
$
\hat{h}(x,y)=
\boxed
{
\begin{matrix}
0\\
2\\
2
\end{matrix}
}(h(x,y))
$. 
Если выделяемое значение равно $1$, то 
$
\hat{h}(x,y)=
\boxed
{
\begin{matrix}
2\\
0\\
2
\end{matrix}
}(h(x,y))
$, а если оно равно $2$, то
$
\hat{h}(x,y)=
\boxed
{
\begin{matrix}
0\\
0\\
2
\end{matrix}
}(h(x,y))
$. 
Если же $h(x,y)$  выделяет некоторое значение, но в основании нет 
соответствующей функции, выберем другое значение, которое на наборах 
этого квадрата принимается нечетное число раз (а такое обязательно найдется), 
и применим соответствующую функцию.

Получившаяся функция $\hat{h}$ 
принимает на наборах квадрата значения $0$ и $2$, 
причем одно из них только в одной вершине. 
Легко видеть, что из такой функции, а также функции вида
$
\boxed
{
\begin{matrix}
2\\
*\\
0
\end{matrix}
}
$ (которая выступает аналогом отрицания на $\{0,2\}$) можно получить искомые 
$m(x,y)$ и $M(x,y)$. А из них уже можно строить неявно полные классы Ореховой. 
В любом случае можно получить одну из следующих неявно полных систем:
$$
\boxed
{
\begin{matrix}
0 & 0 & 0\\
0 & 0 & 0\\
0 & 0 & 2
\end{matrix}
}
\boxed
{
\begin{matrix}
0 & 2 & 2\\
2 & 2 & 2\\
2 & 2 & 2
\end{matrix}
}
\boxed
{
\begin{matrix}
0\\
0\\
0
\end{matrix}
}
\boxed
{
\begin{matrix}
2\\
2\\
2
\end{matrix}
};\;
\boxed
{
\begin{matrix}
0 & 0 & 0\\
0 & 2 & 2\\
0 & 2 & 2
\end{matrix}
}
\boxed
{
\begin{matrix}
0 & 2 & 2\\
2 & 2 & 2\\
2 & 2 & 2
\end{matrix}
}
\boxed
{
\begin{matrix}
0\\
2\\
0
\end{matrix}
}
\boxed
{
\begin{matrix}
2\\
2\\
2
\end{matrix}
};\;
\boxed
{
\begin{matrix}
0 & 0 & 0\\
0 & 0 & 0\\
0 & 0 & 2
\end{matrix}
}
\boxed
{
\begin{matrix}
0 & 0 & 2\\
0 & 0 & 2\\
2 & 2 & 2
\end{matrix}
}
\boxed
{
\begin{matrix}
0\\
0\\
0
\end{matrix}
}
\boxed
{
\begin{matrix}
2\\
0\\
2
\end{matrix}
}.
$$
\end{proof}

\begin{note}
Вместо функции $\phi(x)$, 
которая принимает значения $1$ и $2$ на наборах $0$ и $2$, 
можно потребовать функцию $\hat{\phi}(x)$, 
которая на этих наборах принимает значения $0$ и $1$. 
В таком случае мы сможем свести квадраты, 
у которых левая верхняя вершина $(0,0)$, к квадрату $(0,0)-(2,2)$. 
\end{note}

Используя лемму \ref{kvazi} можно доказать, 
что классы с основаниями 
1.3, 1.4, 2.1, 2.2, 3.3, 3.4 не могут быть неявно предполными.
Действительно, 
так как такие классы не могут совпадать с $\mathfrak{N}$, 
то в них должна быть неквазилинейная функция. 
Но тогда к таким классам применима лемма.

Покажем, что класс с основанием 1.1 
не может быть неявно предполным.
В таком классе должна найтись $f_R\notin R'_2$. Тогда 
\begin{gather*}
f_R(0,1,0,1,2,2,2)=a,\\
f_R(0,1,2,2,0,1,2)=b,\\
f_R(0,1,2,2,2,2,2)=c,
\end{gather*}
причем столбец 
$
\begin{pmatrix}
a\\
b\\
c
\end{pmatrix}
\notin A_1
$. Рассмотрим теперь функцию 
$$
g(x)=
f_R(0,1,
\boxed
{
\begin{matrix}
0\\
2\\
2
\end{matrix}
}(x),
\boxed
{
\begin{matrix}
1\\
2\\
2
\end{matrix}
}(x),
\boxed
{
\begin{matrix}
2\\
0\\
2
\end{matrix}
}(x),
\boxed
{
\begin{matrix}
2\\
1\\
2
\end{matrix}
}(x),
2)=
\boxed
{
\begin{matrix}
a\\
b\\
c
\end{matrix}
}.
$$

С одной стороны, эта функция должна принадлежать исходному классу, с другой, 
столбец значений не совпадает ни с одним столбцов матрицы $A_1$, и $h(x)$ 
не принадлежит основанию своего класса. Противоречие.

\smallskip
Доказательство для основания 3.2 отлично тем, что столбец
$
\begin{pmatrix}
a\\
b\\
c
\end{pmatrix}
$ может быть равен 
$
\boxed
{
\begin{matrix}
1\\
0\\
2
\end{matrix}
}
$. В таком случае рассмотрим 
$$
h(x,y)=
f_R(0,1,x,
\boxed
{
\begin{matrix}
1\\
2\\
2
\end{matrix}
}(x),
y,
\boxed
{
\begin{matrix}
1\\
2\\
2
\end{matrix}
}(y),
2)=
\boxed
{
\begin{matrix}
* & * & 1\\
* & * & *\\
0 & * & 2
\end{matrix}
}.
$$

С одной стороны, 
$
h(0,y)=
\boxed
{
\begin{matrix}
*\\
*\\
1
\end{matrix}
}
$, и так как это должно быть функцией из основания, $h(0,0)=1$.
С другой, 
$
h(x,0)=
\boxed
{
\begin{matrix}
*\\
*\\
0
\end{matrix}
}
$, и так как это должно быть функцией из основания, $h(0,0)=0$.
Противоречие.

\medskip
Развивая это рассуждение, докажем аналогичные утверждения для оснований 
6.1 и 6.2.

В этих случаях из функции $f_R$ можно получить функцию одного из трех видов: 
$$
\boxed
{
\begin{matrix}
* & * & 0\\
* & * & *\\
0 & * & 2
\end{matrix}
};\quad
\boxed
{
\begin{matrix}
* & * & 1\\
* & * & *\\
1 & * & 2
\end{matrix}
};\quad
\boxed
{
\begin{matrix}
* & * & 1\\
* & * & *\\
0 & * & 2
\end{matrix}
}.
$$
По тем же самым причинам третий вариант не реализуется
(он относится только к основанию 6.2), а из первых двух
с помощью функции 
$
\boxed
{
\begin{matrix}
0\\
0\\
2
\end{matrix}
}
$ можно построить функцию 
$
\boxed
{
\begin{matrix}
0 & 0 & 0\\
0 & 0 & 0\\
0 & 0 & 2
\end{matrix}
}
$.

Помимо этого, в основании должна найтись $f_Q\notin Q'_2$. Тогда
\begin{gather*}
f_Q(0,1,0,1,2,2,2)=a',\\
f_Q(0,1,2,2,0,1,2)=b',\\
f_Q(0,1,0,1,0,1,2)=c',
\end{gather*}
причем столбец 
$
\boxed
{
\begin{matrix}
a'\\
b'\\
c'
\end{matrix}
}\notin A_2
$. Построим 
$
h(x,y)=
f_Q(0,1,x,
\boxed
{
\begin{matrix}
1\\
1\\
2
\end{matrix}
}(x),
y,
\boxed
{
\begin{matrix}
1\\
1\\
2
\end{matrix}
}(y),
2)
$, имеющую вид 
$
\boxed
{
\begin{matrix}
c' & * & a'\\
 * & * &  *\\
b' & * &  *\\
\end{matrix}
}
$. Легко видеть, что если хотя бы одно из значений $a',b'$ 
не равно $2$, то подстановкой констант можно получить 
одноместную функцию, не лежащую в основании. 
Поэтому подходят всего два столбца --- 
$
\boxed
{
\begin{matrix}
2\\
2\\
0
\end{matrix}
}
$ и 
$
\boxed
{
\begin{matrix}
2\\
2\\
1
\end{matrix}
}
$. 
Легко видеть, что $h(2,2)=2$, 
в противном случае $h(x,x)$ окажется функцией не из основания.
Теперь из $h(x,y)$ и 
$
\boxed
{
\begin{matrix}
0\\
2\\
2
\end{matrix}
}
$ можно построить функцию 
$
\boxed
{
\begin{matrix}
0 & 2 & 2\\
2 & 2 & 2\\
2 & 2 & 2
\end{matrix}
}
$.

Таким образом, если в классе с основанием 6.1 (6.2) есть 
$f_R\notin R'_2$ и $f_Q\notin Q'_2$, 
то над ним явно выразимы функции
$
\boxed
{
\begin{matrix}
0 & 0 & 0\\
0 & 0 & 0\\
0 & 0 & 2
\end{matrix}
}
\boxed
{
\begin{matrix}
0 & 2 & 2\\
2 & 2 & 2\\
2 & 2 & 2
\end{matrix}
}
\boxed
{
\begin{matrix}
0\\
0\\
0
\end{matrix}
}
\boxed
{
\begin{matrix}
2\\
2\\
2
\end{matrix}
}
$, которые порождают неявно полную систему двойственную $S_1$
относительно $(02)$. 
А значит, такие классы 
не могут быть неявно предполными.

Все основания, подходящие под условие теоремы рассмотрены.
\end{proof}

\subsubsection{Класс L}
\begin{theorem}\label{allL}
Класс линейных функций $L$ есть единственный 
неявно предполный класс, 
основание которого содержит все константы, и не содержится целиком 
ни в одном из классов типа $U$, $M$ или $T_{\epsilon_a,1}$.
\end{theorem}
\begin{proof}
Можно показать, что основание явно замкнутого класса, 
содержащего все 3 константы, 
содержащееся в классе $L$, 
но не содержащееся целиком ни в одном из классов 
типа $U$, $M$ и $T_{\{a\},1}$, 
совпадает с одним из следующих двух оснований 
(или двойственно ему):
$$
\sooo\soiz\siii\sizo\szoz\szzz;\quad
\sooo\soiz\sozi\sioz\siii\sizo\szoz\szio\szzz.
$$

Неявно предполный класс $L$ 
является максимальным надклассом второго основания,
следовательно, 
не существует других неявно предполных классов с таким основанием. 

Так как в первом основании лежат все константы, то вместе с любой 
нелинейной функцией в классе с таким основанием 
по утверждению~\ref{NotIn} должна лежать и 
одноместная нелинейная функция. 
Но в основании таких функций нет, а значит, любой класс с таким 
основанием содержится в неявно предполном классе $L$.
\end{proof}

\subsubsection{Класс Слупецкого}
\begin{theorem} \label{allSlup}
Класс квазилинейных функций $\mathfrak{N}$ есть единственный 
неявно предполный класс, 
основание которого содержит все константы, 
и не содержится целиком ни в одном 
из предполных по суперпозиции классов, 
за исключением класса Слупецкого.
\end{theorem}

\begin{proof}

Выясним, какими могут быть основания $\mathfrak{A}$, 
удовлетворяющие условию теоремы.

В доказательстве теоремы о функциональной полноте в $P_3$ Яблонский 
из условия невложимости класса в классы типа $U$ получает 6 систем 
одноместных функций таких, 
что исходный класс должен содержать хотя бы одну из них. 
Рассмотрим каждую из этих систем и добавим условие невложимости 
в класс $L$ и классы типа $M$ и $T_{\epsilon_a,1}$.

1) 
$
\boxed
{
\begin{matrix}
0\\
0\\
2
\end{matrix}
}
\boxed
{
\begin{matrix}
0\\
2\\
1
\end{matrix}
}\in \mathfrak{A}
$.

Сформулируем лемму:
\begin{lemma}
\cite{Yablonskiy} Если $f(\tilde{x})\notin T_{\{0\},1}$, 
то с помощью констант и функции
$
\boxed
{
\begin{matrix}
0\\
2\\
1
\end{matrix}
}
$ можно построить $g(x)$ такую, что $g(0)\neq 0$ и $g(0)\neq g(1)$.
\end{lemma}

Все условия леммы выполнены, поэтому в основании должна лежать такая 
функция $g(x)$. Очевидно, что с помощью функции 
$
\boxed
{
\begin{matrix}
0\\
2\\
1
\end{matrix}
}
$ из нее можно получить функцию вида 
$
\boxed
{
\begin{matrix}
1\\
0\\
*
\end{matrix}
}
$ или функцию вида 
$
\boxed
{
\begin{matrix}
1\\
2\\
*
\end{matrix}
}
$. 

Можно показать, что в каждом из этих случаев мы 
сможем воспользоваться леммой \ref{kvazi}.

С учетом замечания к лемме \ref{kvazi}, мы выразили все
необходимые функции. 
Таким образом, класс с таким или более широким основанием,
содержащий неквазилинейную функцию, неявно полон.

Можно показать, 
что все оставшиеся случаи можно свести к функции 
$
\boxed
{
\begin{matrix}
1\\
0\\
1
\end{matrix}
}
$.
\medskip

2)
$
\boxed
{
\begin{matrix}
0\\
1\\
0
\end{matrix}
}
\boxed
{
\begin{matrix}
0\\
2\\
2
\end{matrix}
}
\boxed
{
\begin{matrix}
1\\
1\\
2
\end{matrix}
}\in \mathfrak{A}
$.
Покажем, 
что уже этих функций достаточно для того, 
чтобы воспользоваться леммой \ref{kvazi}.
$$
\boxed
{
\begin{matrix}
0\\
1\\
0
\end{matrix}
}
\left(
\boxed
{
\begin{matrix}
1\\
1\\
2
\end{matrix}
}
\right)=
\boxed
{
\begin{matrix}
1\\
1\\
0
\end{matrix}
};\quad
\boxed
{
\begin{matrix}
0\\
2\\
2
\end{matrix}
}
\left(
\boxed
{
\begin{matrix}
1\\
1\\
0
\end{matrix}
}
\right)=
\boxed
{
\begin{matrix}
2\\
2\\
0
\end{matrix}
};\quad
\boxed
{
\begin{matrix}
2\\
2\\
0
\end{matrix}
}
\left(
\boxed
{
\begin{matrix}
2\\
2\\
0
\end{matrix}
}
\right)=
\boxed
{
\begin{matrix}
0\\
0\\
2
\end{matrix}
}.
$$

Таким образом, класс, содержащий эти функции, 
а также неквазилинейную функцию, 
неявно предполон.

3)
$
\boxed
{
\begin{matrix}
0\\
0\\
1
\end{matrix}
}
\boxed
{
\begin{matrix}
0\\
2\\
0
\end{matrix}
}\in \mathfrak{A}
$. 
Такое основание содержится в классе $T_{\{0\},1}$, 
а значит у нас должна быть функция не из этого класса. 
То есть одна из
$$
\boxed
{
\begin{matrix}
1\\
0\\
2
\end{matrix}
}
\boxed
{
\begin{matrix}
1\\
1\\
2
\end{matrix}
}
\boxed
{
\begin{matrix}
1\\
2\\
0
\end{matrix}
}
\boxed
{
\begin{matrix}
1\\
2\\
1
\end{matrix}
}
\boxed
{
\begin{matrix}
1\\
2\\
2
\end{matrix}
}
\boxed
{
\begin{matrix}
2\\
0\\
1
\end{matrix}
}
\boxed
{
\begin{matrix}
2\\
1\\
0
\end{matrix}
}
\boxed
{
\begin{matrix}
2\\
1\\
1
\end{matrix}
}
\boxed
{
\begin{matrix}
2\\
1\\
2
\end{matrix}
}
\boxed
{
\begin{matrix}
2\\
2\\
1
\end{matrix}
}
$$

Можно показать, во-первых, что при добавлении функции 
$
\boxed
{
\begin{matrix}
2\\
1\\
2
\end{matrix}
}
$ мы можем воспользоваться леммой~\ref{kvazi}, а во-вторых, 
при добавлении любой другой функции из списка, можно выразить 
$
\boxed
{
\begin{matrix}
2\\
1\\
2
\end{matrix}
}
$.

4)
$
\boxed
{
\begin{matrix}
0\\
0\\
2
\end{matrix}
}
\boxed
{
\begin{matrix}
0\\
1\\
1
\end{matrix}
}
\boxed
{
\begin{matrix}
0\\
2\\
2
\end{matrix}
}\in \mathfrak{A}
$. Такое основание содержится сразу в двух предполных по суперпозиции классах: 
$T_{\{0\},1}$ и $M_1$. Чтобы "выпасть" из $T_{\{0\},1}$,
основание должно содержать одну из 10 функций не из 
этого класса. 
Несложно показать, 
что добавление любой из этих функций можно свести 
к функции
$
\boxed
{
\begin{matrix}
1\\
1\\
2
\end{matrix}
}
$. 

Поэтому без ограничения общности можно считать, 
что в основании лежит функция 
$
\boxed
{
\begin{matrix}
1\\
1\\
2
\end{matrix}
}\notin T_{\{0\},1}
$. 
Однако такое основание все еще содержится в классе $M_1$, 
а значит, должно содержать немонотонную функцию. 
То есть одну из функций
$$
\boxed
{
\begin{matrix}
0\\
1\\
0
\end{matrix}
}
\boxed
{
\begin{matrix}
0\\
2\\
0
\end{matrix}
}
\boxed
{
\begin{matrix}
0\\
2\\
1
\end{matrix}
}
\boxed
{
\begin{matrix}
1\\
0\\
0
\end{matrix}
}
\boxed
{
\begin{matrix}
1\\
0\\
1
\end{matrix}
}
\boxed
{
\begin{matrix}
1\\
0\\
2
\end{matrix}
}
\boxed
{
\begin{matrix}
1\\
1\\
0
\end{matrix}
}
\boxed
{
\begin{matrix}
1\\
2\\
0
\end{matrix}
}
\boxed
{
\begin{matrix}
1\\
2\\
1
\end{matrix}
}
\boxed
{
\begin{matrix}
2\\
0\\
0
\end{matrix}
}
\boxed
{
\begin{matrix}
2\\
0\\
1
\end{matrix}
}
\boxed
{
\begin{matrix}
2\\
0\\
2
\end{matrix}
}
\boxed
{
\begin{matrix}
2\\
1\\
0
\end{matrix}
}
\boxed
{
\begin{matrix}
2\\
1\\
1
\end{matrix}
}
\boxed
{
\begin{matrix}
2\\
1\\
2
\end{matrix}
}
\boxed
{
\begin{matrix}
2\\
2\\
0
\end{matrix}
}
\boxed
{
\begin{matrix}
2\\
2\\
1
\end{matrix}
}
$$
Применяя к этим функциям одну из функций 
$
\boxed
{
\begin{matrix}
0\\
0\\
2
\end{matrix}
}
\boxed
{
\begin{matrix}
0\\
2\\
2
\end{matrix}
}
$, можно свести все к четырем функциям:
$$
\boxed
{
\begin{matrix}
0\\
2\\
0
\end{matrix}
}
\boxed
{
\begin{matrix}
2\\
0\\
0
\end{matrix}
}
\boxed
{
\begin{matrix}
2\\
0\\
2
\end{matrix}
}
\boxed
{
\begin{matrix}
2\\
2\\
0
\end{matrix}
}
$$
Любой из функций 
$
\boxed
{
\begin{matrix}
2\\
0\\
0
\end{matrix}
}
\boxed
{
\begin{matrix}
2\\
2\\
0
\end{matrix}
}
$ достаточно для того, 
чтобы можно было воспользоваться леммой \ref{kvazi}.
А в силу
$$
\boxed
{
\begin{matrix}
0\\
2\\
0
\end{matrix}
}
\left(
\boxed
{
\begin{matrix}
1\\
1\\
2
\end{matrix}
}
\right)=
\boxed
{
\begin{matrix}
2\\
2\\
0
\end{matrix}
};\quad
\boxed
{
\begin{matrix}
2\\
0\\
2
\end{matrix}
}
\left(
\boxed
{
\begin{matrix}
0\\
1\\
1
\end{matrix}
}
\right)=
\boxed
{
\begin{matrix}
2\\
0\\
0
\end{matrix}
}
$$
мы сможем ей воспользоваться в любом из случаев.

Таким образом, класс, содержащий неквазилинейную функцию, а также 
$
\boxed
{
\begin{matrix}
0\\
0\\
2
\end{matrix}
}
\boxed
{
\begin{matrix}
0\\
1\\
1
\end{matrix}
}
\boxed
{
\begin{matrix}
0\\
2\\
2
\end{matrix}
}
$, неявно полон.

5)
$
\boxed
{
\begin{matrix}
0\\
0\\
1
\end{matrix}
}
\boxed
{
\begin{matrix}
1\\
2\\
2
\end{matrix}
}\in \mathfrak{A}
$. Эти функции содержатся в классе $T_{\{1\},1}$, а значит 
в основании должна лежать хотя бы одна из функций
$$
\boxed
{
\begin{matrix}
0\\
0\\
2
\end{matrix}
}
\boxed
{
\begin{matrix}
0\\
2\\
0
\end{matrix}
}
\boxed
{
\begin{matrix}
0\\
2\\
1
\end{matrix}
}
\boxed
{
\begin{matrix}
0\\
2\\
2
\end{matrix}
}
\boxed
{
\begin{matrix}
1\\
0\\
2
\end{matrix}
}
\boxed
{
\begin{matrix}
1\\
2\\
0
\end{matrix}
}
\boxed
{
\begin{matrix}
2\\
0\\
0
\end{matrix}
}
\boxed
{
\begin{matrix}
2\\
0\\
1
\end{matrix}
}
\boxed
{
\begin{matrix}
2\\
0\\
2
\end{matrix}
}
\boxed
{
\begin{matrix}
2\\
2\\
0
\end{matrix}
}.
$$
Можно показать, 
что вместе с любой из этих функций в основании будет лежать функция
$
\boxed
{
\begin{matrix}
0\\
0\\
2
\end{matrix}
}
$.
В силу соотношений
$$
\boxed
{
\begin{matrix}
0\\
0\\
2
\end{matrix}
}
\left(
\boxed
{
\begin{matrix}
1\\
2\\
2
\end{matrix}
}
\right)=
\boxed
{
\begin{matrix}
0\\
2\\
2
\end{matrix}
};\quad
\boxed
{
\begin{matrix}
0\\
0\\
1
\end{matrix}
}
\left(
\boxed
{
\begin{matrix}
0\\
2\\
2
\end{matrix}
}
\right)=
\boxed
{
\begin{matrix}
0\\
1\\
1
\end{matrix}
}
$$
можно считать, 
что основание $\mathfrak{A}$ содержит в себе 
все функции из пункта 4). 
А значит и класс с таким основанием неявно полон.

6)
$
\boxed
{
\begin{matrix}
1\\
2\\
0
\end{matrix}
}\in \mathfrak{A}
$. 
Так как всякая нелинейная одноместная функция 
выпускает ровно одно значение,
то можно применить лемму \ref{nonlin}, тем самым получив все 
одноместные функции, выпускающие хотя бы одно значение. 
Воспользовавшись леммой \ref{kvazi} получим, что класс, содержащий 
циклическую перестановку, нелинейную функцию, а также 
неквазилинейную функцию, неявно полон.

\medskip
Все основания, удовлетворяющие условию теоремы, рассмотрены. Теорема доказана.
\end{proof}

\subsection{Основные результаты}
\begin{theorem}\label{Main}
Система функций трехзначной логики полна в $P_3$ 
по неявной выразимости тогда и только тогда, 
когда она целиком не содержится ни в одном из следующих 54 неявно предполных классов:
\begin{gather*}
S, L, T_{\{0\},0}, T_{\{1\},0}, T_{\{2\},0},
W_0, W_1, W_2, Y_0, Y_1, Y_2,\\
\Sigma^{\{0,1\}}_L, \Sigma^{\{0,1\}}_S, \Sigma^{\{0,1\}}_K, \Sigma^{\{0,1\}}_D,
\Sigma^{\{0,2\}}_L, \Sigma^{\{0,2\}}_S, \Sigma^{\{0,2\}}_K, \Sigma^{\{0,2\}}_D,
\Sigma^{\{1,2\}}_L, \Sigma^{\{1,2\}}_S, \Sigma^{\{1,2\}}_K, \Sigma^{\{1,2\}}_D,\\
\Sigma^{\{0,1\}\{2\}}_L, 
\Sigma^{\{0,1\}\{2\}}_{T_0}, \Sigma^{\{0,1\}\{2\}}_{T_1},
\Sigma^{\{0,1\}\{2\}}_K, \Sigma^{\{0,1\}\{2\}}_D,
\Sigma^{\{0,2\}\{1\}}_L, 
\Sigma^{\{0,2\}\{1\}}_{T_0}, \Sigma^{\{0,2\}\{1\}}_{T_1},\\
\Sigma^{\{0,2\}\{1\}}_K, \Sigma^{\{0,2\}\{1\}}_D,
\Sigma^{\{1,2\}\{0\}}_L, 
\Sigma^{\{1,2\}\{0\}}_{T_0}, \Sigma^{\{1,2\}\{0\}}_{T_1},
\Sigma^{\{1,2\}\{0\}}_K, \Sigma^{\{1,2\}\{0\}}_D,\\
KM_1, DM_1, KM_2, DM_2, KM_3, DM_3, 
R'_0, R'_1, R'_2, Q'_0, Q'_1, Q'_2, F'_0, F'_1, F'_2, 
\mathfrak{N}.
\end{gather*}
\end{theorem}

\begin{proof}
Во второй главе было доказано,
что каждый из 54 классов,
указанных в теореме, является неявно предполным.
Следовательно, 
если система целиком содержится в одном из этих классов, 
то она является неявно неполной. 

В работе~\cite{OMKZ_Pk} было доказано, 
что любая неявно неполная система функций в $P_k$ 
содержится в некотором неявно предполном классе.
Совокупность теорем \ref{all0const}, 
\ref{all1const}, \ref{all2const}, 
\ref{allU}, \ref{allM}, \ref{allT1}, 
\ref{allL}, \ref{allSlup} доказывает, 
что каким бы ни было основание, 
класс с таким основанием может быть 
неявно предполным только в том случае, 
если он совпадает с одним из описанных 
неявно предполных классов.
Следовательно, 
если система не содержится ни в одном из указанных классов, 
то она является неявно полной.

Тем самым доказан критерий неявной полноты в трехзначной логике
в терминах предполных классов.
\end{proof}

Система $\mathfrak{A}$ называется \textit{слабо неявно полной},
если система $\mathfrak{A}\cup \{0,1,\dots,k-1\}$ неявно полна.

\begin{theorem}\label{Main_weak}
Система функций $\mathfrak{A}$ трехзначной логики 
слабо неявно полна в $P_3$ тогда и только тогда, 
когда она целиком не содержится ни в одном из 
следующих 28 неявно предполных классов:
\begin{gather*}
W_0, W_1,W_2,
\Sigma^{\{0,1\}\{2\}}_L, 
\Sigma^{\{0,1\}\{2\}}_K, \Sigma^{\{0,1\}\{2\}}_D,
\Sigma^{\{0,2\}\{1\}}_L,\\ 
\Sigma^{\{0,2\}\{1\}}_K, \Sigma^{\{0,2\}\{1\}}_D,
\Sigma^{\{1,2\}\{0\}}_L, 
\Sigma^{\{1,2\}\{0\}}_K, \Sigma^{\{1,2\}\{0\}}_D,\\
KM_1, DM_1, KM_2, DM_2, KM_3, DM_3, 
R'_0, R'_1, R'_2, Q'_0, Q'_1, Q'_2, F'_0, F'_1, F'_2, 
\mathfrak{N}.
\end{gather*}
\end{theorem}

\begin{proof}
Заметим, что приведенные классы~--- 
в точности те неявно предполные классы, 
которые содержат все константы.
Очевидно, что добавление к этим классам констант не расширяет их,
и они являются слабо неявно неполными.
Следовательно, 
всякая подсистема любого из этих классов слабо неявно неполна.

Пусть $\mathfrak{A}$ не содержится ни в одном 
из указанных 28 классов.
Тогда система $\mathfrak{A}\cup\{0,1, 2\}$ 
не содержится ни в одном из 54 неявно предполных классов 
и по теореме~\ref{Main} является неявно полной.
А значит, система $\mathfrak{A}$ является слабо неявно полной.
\end{proof}


\begin{thebibliography}{40}


\bibitem{OMKZ_P2} Касим-Заде О. М. 
О неявной выразимости булевых функций //
Вестник МГУ. Серия 1. Математика. Механика.~--- 
1995.~--- №2.~--- С. 44--49.

\bibitem{OMKZ_crypto} Касим-Заде О. М.
О неявной выразимости в двузначной логике и криптоизоморфизмах 
двухэлементных алгебр~//
Доклады РАН.~--- 1996.~--- Т. 348, №~3.~--- С. 299--301.

\bibitem{OMKZ_bf_param} Касим-Заде О. М. 
О сложности параметрических представлений булевых функций //
Математические вопросы кибернетики.~--- 
1998.~--- Вып. 7.~--- С. 85--160.

\bibitem{OMKZ_Pk} Касим-Заде О. М. 
О неявной полноте в $k$-значной логике //
Вестник МГУ. Серия 1. Математика. Механика.~--- 
2007.~--- №3.~--- С. 9--13.


\bibitem{Kuznetsov} Кузнецов А. В. 
О средствах для обнаружения невыводимости или невыразимости //
Логический вывод.~--- М.: Наука, 1979.~--- С. 5--33.

\bibitem{Orehova_Pk} Орехова Е. А. Об одном критерии неявной полноты 
в $k$-значной логике //
Математические вопросы кибернетики.~--- 
2002.~--- Вып. 11.~--- С. 77--90.

\bibitem{Orehova_P3} Орехова Е. А. Об одном критерии неявной полноты 
в трёхзначной логике //
Математические вопросы кибернетики.~--- 
2003.~--- Вып. 12.~--- С. 27--74.


\bibitem{Starostin_T} Старостин М. В.
О некоторых неявно предполных классах функций, 
сохраняющих подмножества //
Вестник МГУ. Серия 1. Математика. Механика.~--- 
2018.~--- №6.~--- С. 36--40.

\bibitem{Starostin_M} Старостин М. В.
О некоторых неявно предполных классах монотонных функций в $P_k$ //
Дискретная математика.~--- 2018.~--- Т. 30, №~4.~--- С. 106–-114.


\bibitem{Ugolnikov} Угольников А. Б. Классы Поста.
М.: Изд.-во ЦПИ при мех.-мат. факультете МГУ, 2008.


\bibitem{Yablonskiy} Яблонский С. В. Функциональные построения 
в $k$-значной логике //
Труды математического института АН СССР им. Стеклова.
~--- М.: Изд-во АН СССР, 1958.~--- С. 5--142.


\bibitem{Lau} Lau D. Function Algebras on Finite Sets.~---
Springer, 2006.





\end{thebibliography}
\end{document}